\documentclass[usenatbib]{mn2e}

\usepackage{graphicx}
\usepackage{amssymb}
\usepackage{rotating}

\newcommand{\apj}{ApJ}
\newcommand{\apjl}{ApJ}
\newcommand{\aap}{A\&A}
\newcommand{\mnras}{MNRAS}
\newcommand{\nat}{Nature}
\newcommand{\gcn}{GCN Circular}

\title[Testing the blast wave model with \emph{Swift} GRBs]
{Testing the blast wave model with \emph{Swift} GRBs}

\author[P.A.~Curran et al.]
{P.A.~Curran$^{1,2}$\thanks{e-mail: pac@mssl.ucl.ac.uk}, 
R.L.C.~Starling$^3$,
A.J.~van~der~Horst$^{4}$,
R.A.M.J.~Wijers$^1$ \\ 
$^1$Astronomical Institute, University of Amsterdam, Kruislaan 403, 1098\,SJ Amsterdam, The Netherlands \\
$^2$Mullard Space Science Laboratory, University College of London, Holmbury St Mary, Dorking, Surrey RH5\,6NT, UK\\ 
$^3$Department~of~Physics~and~Astronomy, University~of~Leicester, University~Road, Leicester~LE1~7RH, UK \\
$^4$NASA Postdoctoral Program Fellow, NSSTC, 320 Sparkman Drive, Huntsville, AL 35805, USA \\
}

\begin{document}

\date{Accepted 2009 January 29. Received 2009 January 27; in original form 2008 December 15}

\pagerange{\pageref{firstpage}--\pageref{lastpage}} \pubyear{}

\maketitle

\label{firstpage}


\begin{abstract}
%
The complex structure of the light curves of \emph{Swift} GRBs has made the identification of breaks, and the interpretation of the blast wave caused by the burst, more difficult than in the pre-\emph{Swift} era.
We aim to identify breaks, which are possibly hidden, and to constrain the blast wave parameters; electron energy distribution, $p$, density profile of the circumburst medium, $k$, and the continued energy injection index, $q$. 
We do so by comparing the observed multi-wavelength light curves and X-ray spectra of our sample to the predictions of the blast wave model. 
We can successfully interpret all of the bursts in our sample of 10, except two, within this framework and we can estimate, with confidence, the electron energy distribution index for 6 of the sample. Furthermore  we identify jet breaks in a number of the bursts. 
A statistical analysis of the distribution of $p$ reveals that, even in the most conservative case of least scatter, the values are not consistent with a single, universal value.  The values of $k$ suggest that the circumburst density profiles are not drawn from only one of the constant density or wind-like media populations. 
\end{abstract}



\section{Introduction}\label{breaks:introduction}

The afterglow emission of Gamma-Ray Bursts (GRBs) is generally well described by the blast wave model \citep{rees1992:MNRAS258,meszaros1997:ApJ476,meszaros1998:ApJ499}. This model details the temporal and spectral behaviour of the emission that is created by external shocks when a collimated ultra-relativistic jet ploughs into the circumburst medium, driving a blast wave ahead of it. 
The level of collimation, or jet opening angle, has important implications for the energetics of the underlying physical process, progenitor models, and the possible use of GRBs as standard candles. The signature of the collimation, according to simple analytical models, is an achromatic temporal steepening  or `jet break' at approximately one day in an otherwise decaying, power-law light curve; from the time of this break, the jet opening angle can be estimated \citep{rhoads1997:ApJ487}.

Since the launch of the \emph{Swift} satellite \citep{gehrels2004:ApJ611} it has become clear that this model for GRBs cannot, in its current form, explain the full complexity of observed light curve features and the lack of observed achromatic temporal breaks. The unexpected features detected, such as steep decays, plateau phases (e.g., \citealt{tagliaferri2005:Natur436,nousek2006:ApJ642,obrien2006:ApJ647}) and a large number of X-ray flares (e.g., \citealt{burrows2007:RSPT365,chincarini2007:ApJ671,falcone2007:ApJ671}) have revealed the complexity of these sources up to about one day since the initial event, which is yet to be fully understood. These superimposed features also make it difficult to measure the underlying power-law features on which the blast wave model is based, and may lead to misinterpretations of the afterglows. 

Achromatic temporal breaks are not observed in the majority of bursts, up to weeks in a few bursts (e.g., \citealt{panaitescu2006:MNRAS369,burrows2007:astro.ph2633}). In some afterglows, a break is unobserved in both the X-ray and optical light curves, while in other bursts a break is observed in one regime but not the other (e.g., \citealt{liang2008:ApJ675}). Previous comprehensive studies of \emph{Swift} light curves in multiple wavelengths \citep{liang2007:ApJ670,liang2008:ApJ675} have attempted to identify achromatic temporal breaks due either to jet breaks, or the cessation of continued energy injection (i.e. end of the plateau phase), mainly as they concern the energy requirement of bursts. 

Identification of jet breaks is also important as a test of the blast wave physics as it represents the time at which sideways spreading of the jet becomes important and the edges of the jet become visible. The post jet break phase also gives a direct measurement of the electron energy distribution, $p$, as the temporal decay, at all wavelengths, during this phase should have a value equal to that of $p$. However, due to differences between broken power-law fits and the underlying parameters \citep{johannesson2006:ApJ640}, this method may systematically underestimate the true value of $p$. The value of $p$ may also be measured from the spectral information of the optical or X-ray bands; the accuracy of this method is then dependent on the quality of the spectra used. A study of \emph{BeppoSAX} bursts used broadband optical to X-ray Spectral Energy Distributions (SEDs) to accurately constrain the value of $p$ via the spectral and temporal parameters of both bands, and a comparison to the blast wave model \citep{starling2008:ApJ672}. The same authors also use this method to estimate the density profile of the circumburst medium, $k$, as the temporal slope may depend on this value as defined by $\rho \propto r^{-k}$; a constant density medium, $k=0$ causing a more shallow decay than a wind like medium, $k=2$. Similar interpretations of the temporal and spectral information may also be used to estimate the effect, if any, that continued energy injection has on the light curves. Continued energy injection is used to explain the shallow decay phase observed in many \emph{Swift} GRBs and is usually assumed to take the form of $E \propto t^{q}$ \citep{nousek2006:ApJ642}. The additional energy can be due to slower shells of matter catching up with to blast wave, or a continued activity of the central engine. In either case, the addition of the extra energy to the blast wave causes the flux to drop off less rapidly than it would otherwise, and is observed as a shallower than expected decay. The cessation of the energy injection is marked by a temporal break to the normal decay phase as predicted by the blast wave model which implies, if it is due to continued activity of the central engine, that the engine must be active for up to about a day. Such late activity of the central engine has been observed in the form of episodic flares  \citep{curran2008:A&A487} but such late, smooth and continuous activity has yet to be confirmed.

In this paper we interpret a sample of afterglows of \emph{Swift} GRBs, to constrain the blast wave parameters: electron energy distribution, $p$, density profile of the circumburst medium, $k$, and the continued energy injection index, $q$. 
In \S\ref{breaks:observations} we introduce our sample and the method of reduction, while in \S\ref{breaks:results} we present the results of our temporal and spectral analysis. In \S\ref{breaks:discussion} we interpret the results in the context of the blast wave model and discuss their implications in the overall context of GRB observations. We summarise our findings in \S\ref{breaks:conclusion}.
Throughout, we use the convention that a power-law flux is given as $F \propto t^{-\alpha} \nu^{-\beta}$ where $\alpha$ is the temporal decay index and $\beta$ is the spectral index.  All uncertainties are quoted at the $1\sigma$ confidence level.


\section{Observations and analyses}\label{breaks:observations}

\subsection{Sample selection}\label{breaks:sample}

The bursts in our sample were chosen from an inspection of previous literature on samples of temporal breaks \citep{liang2007:ApJ670,liang2008:ApJ675}. Bursts were also chosen from a comparison of the pre-reduced \emph{Swift} X-ray Telescope (XRT, \citealt{burrows2005:SSRv120}) light curves in the on-line repository \citep{evans2007:A&A469} up to the end of February  2008, and from the literature of optical data.

Our sample (Table \ref{breaks:sample-table}) consists of bursts with X-ray and optical light curves with good enough time coverage to allow for the underlying single power-law, or broken power-law, to be determined. The bursts are also well sampled enough in the X-ray and optical to constrain the spectral indices, $\beta_{{\rm X}}$ and $\beta_{{\rm opt}}$. Bursts with published optical data were preferred over bursts with only GCN (Gamma-ray bursts Coordinates Network Circular) data which, by their very nature, are preliminary and hence difficult to compare with other data sets. However, in the cases of bursts with significant amounts of data from a single source, which we assume to be self-consistent, these data were included even if published only in GCNs. We did not confine our sample to bursts with clear breaks in either the X-ray or optical bands as we wanted to include the possibility of hidden or not very obvious breaks, particularly in the X-ray \citep{curran2008:MNRAS386}, or late, unobserved breaks.

\begin{figure*}
 \centering 
\begin{minipage}{80mm}
\resizebox{80mm}{!}{\includegraphics[angle=-90]{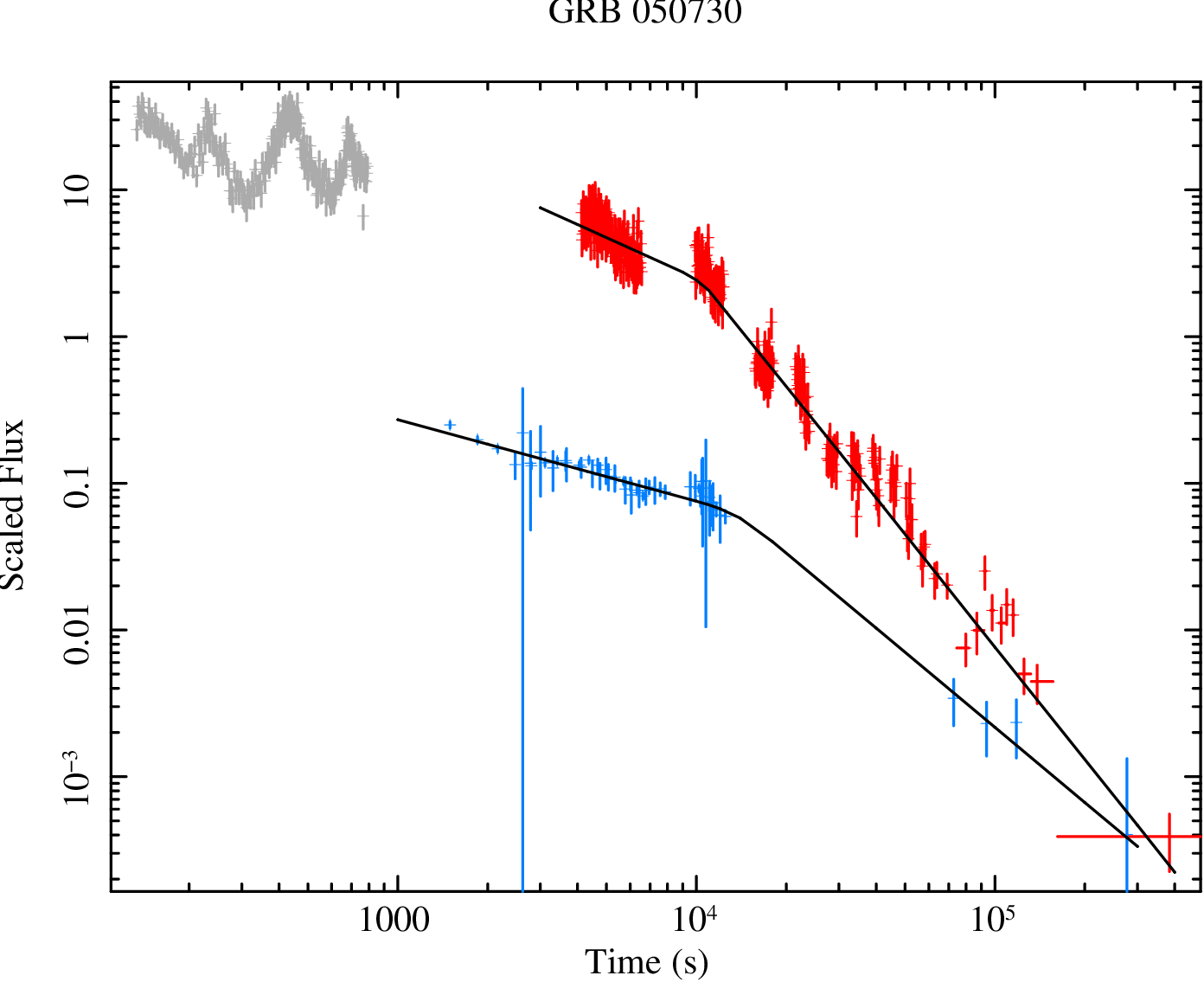} }
\end{minipage}
\hspace{3mm}
\vspace{3mm}
\begin{minipage}{80mm}
\resizebox{80mm}{!}{\includegraphics[angle=-90]{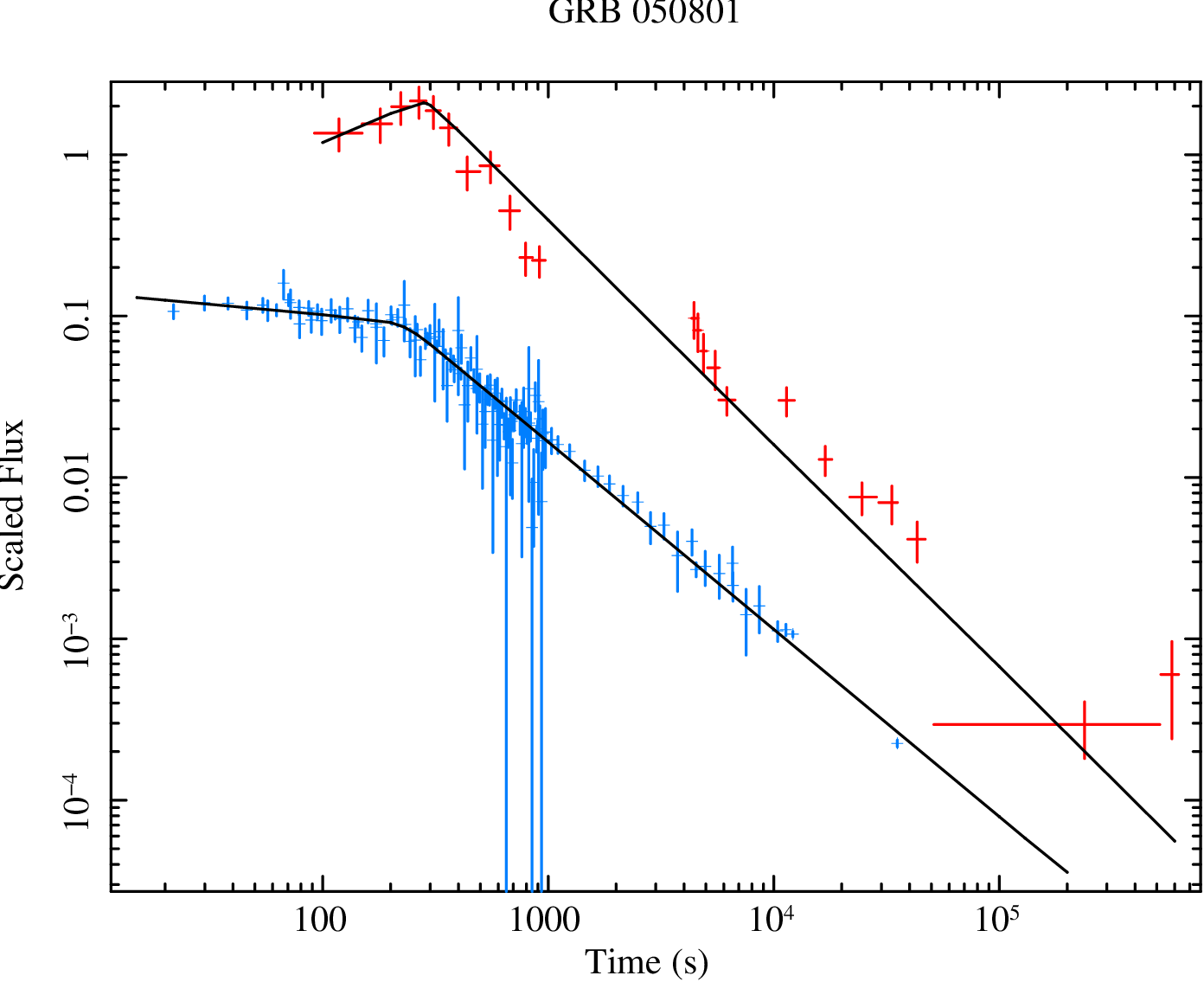} }
\end{minipage}
\vspace{3mm}
\begin{minipage}{80mm}
\resizebox{80mm}{!}{\includegraphics[angle=-90]{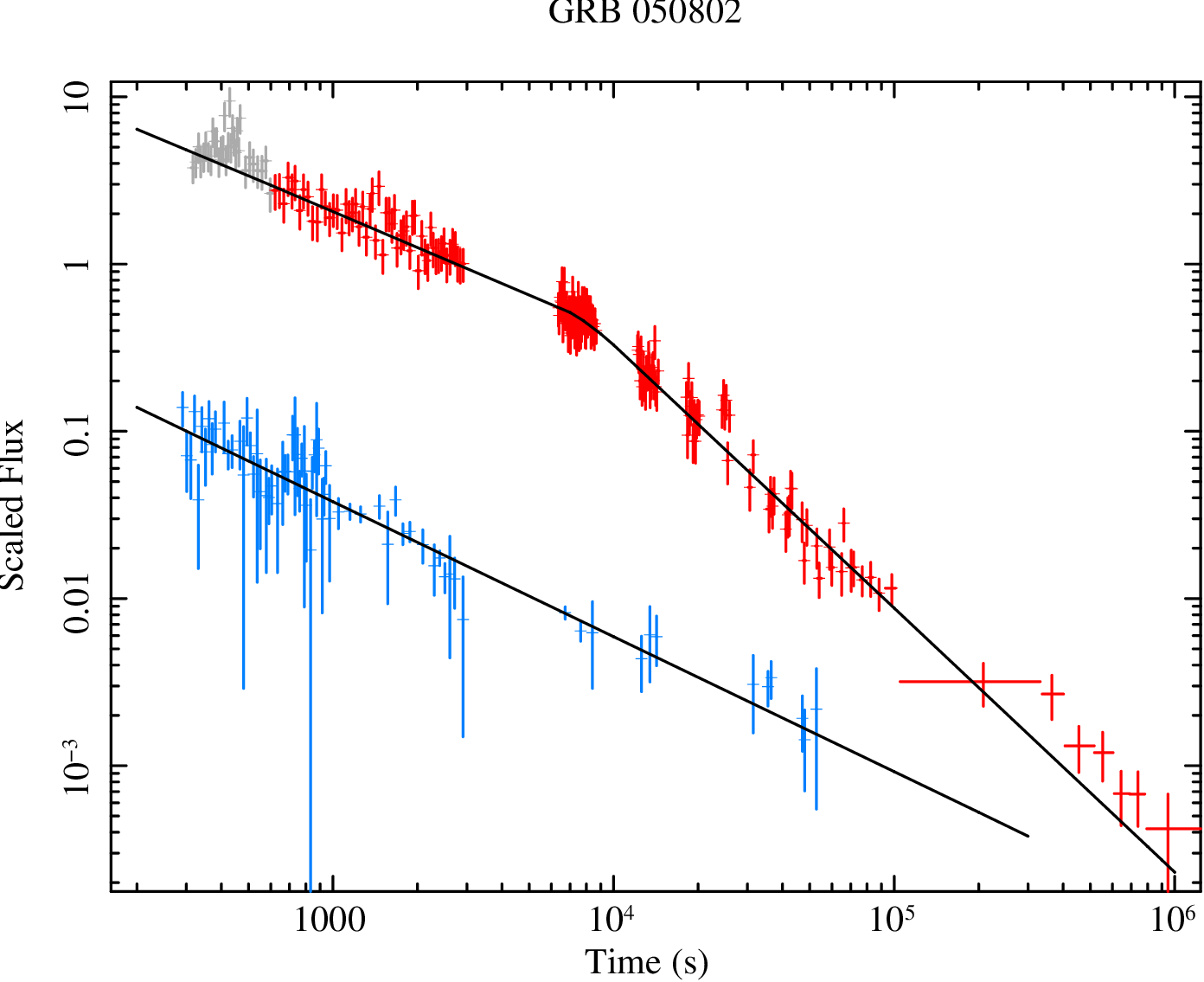} }
\end{minipage}
\hspace{3mm}
\begin{minipage}{80mm}
\resizebox{80mm}{!}{\includegraphics[angle=-90]{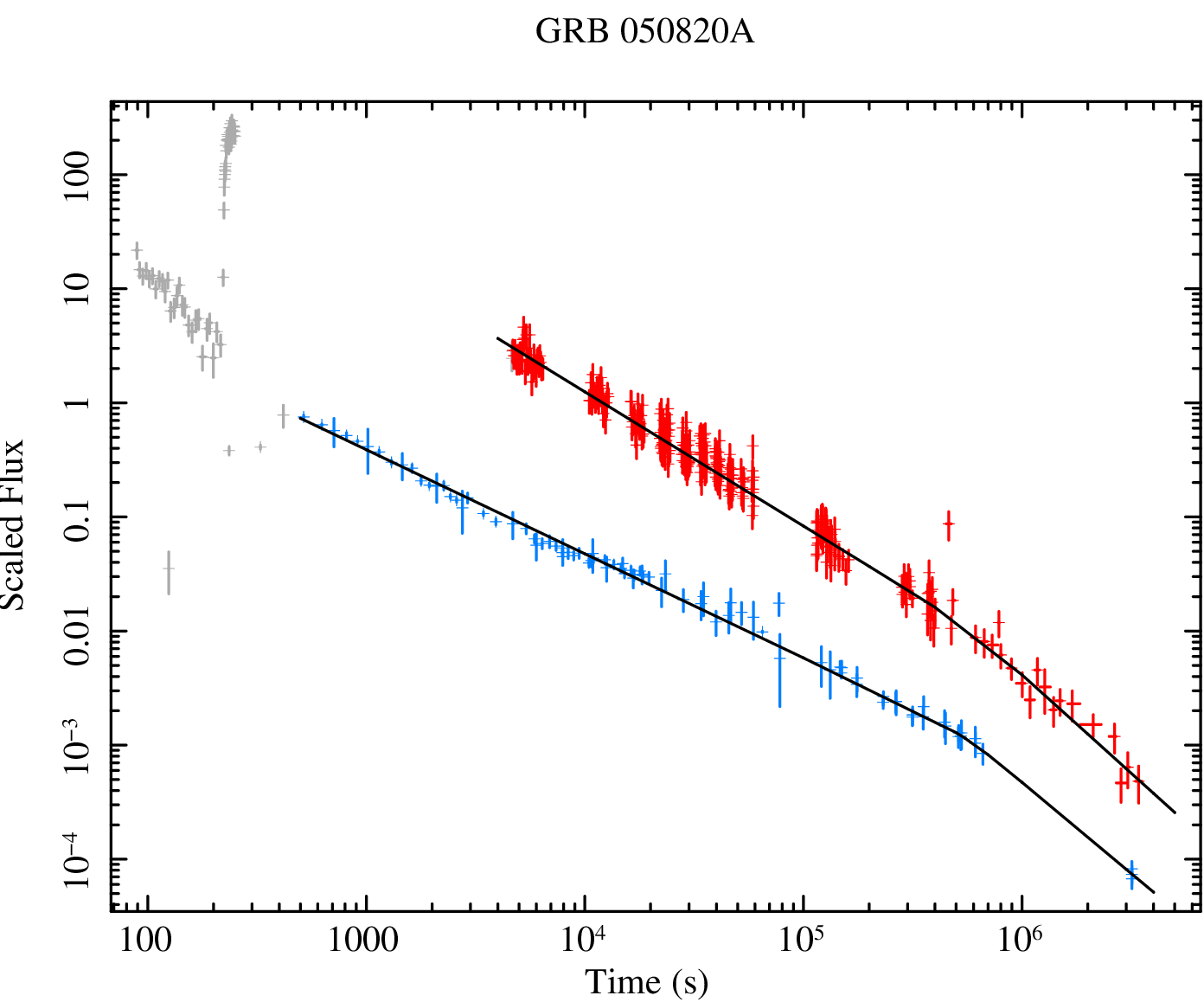} }
\end{minipage}
\vspace{3mm}
\begin{minipage}{80mm}
\resizebox{80mm}{!}{\includegraphics[angle=-90]{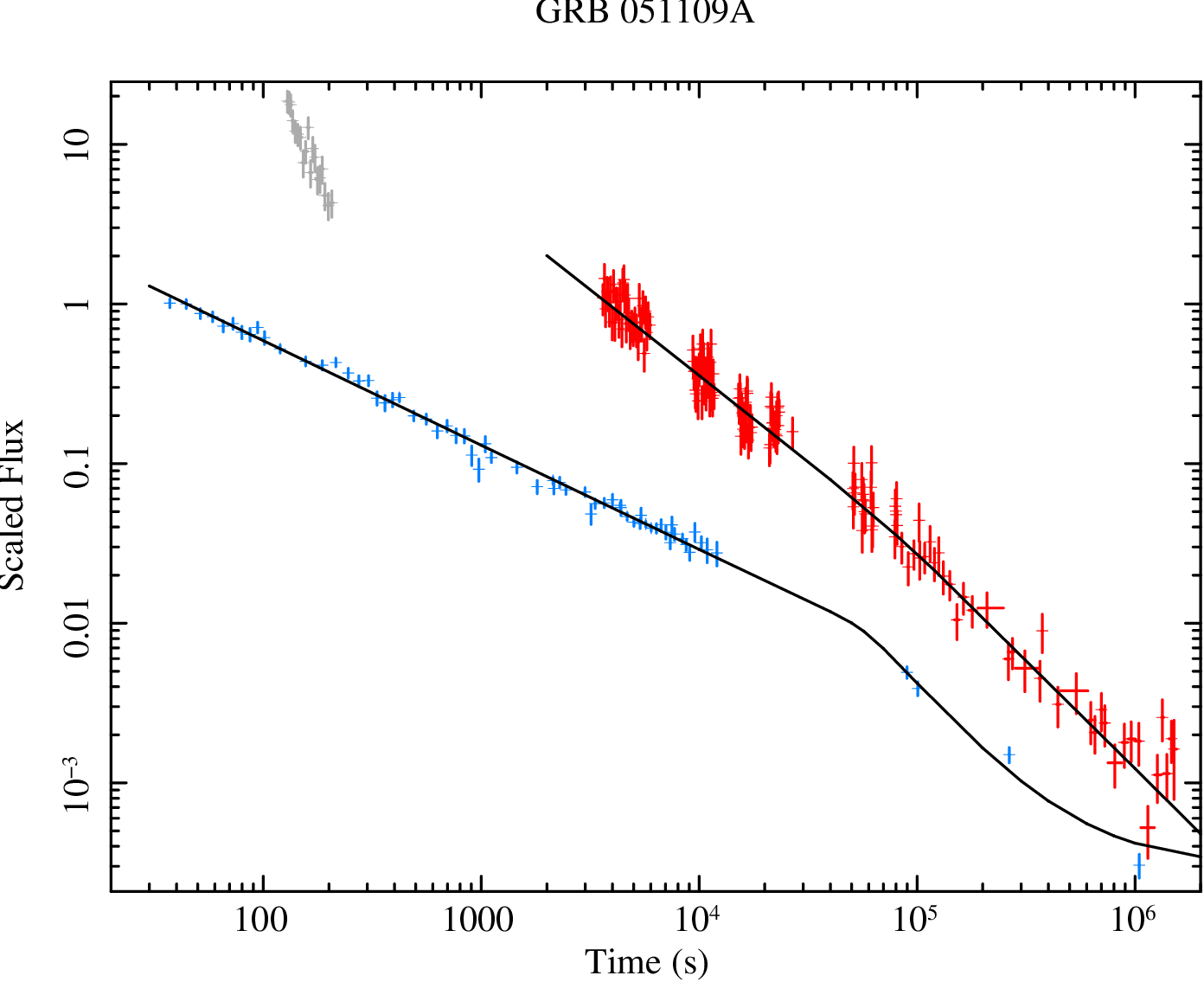} }
\end{minipage}
\hspace{3mm}
\begin{minipage}{80mm}
\resizebox{80mm}{!}{\includegraphics[angle=-90]{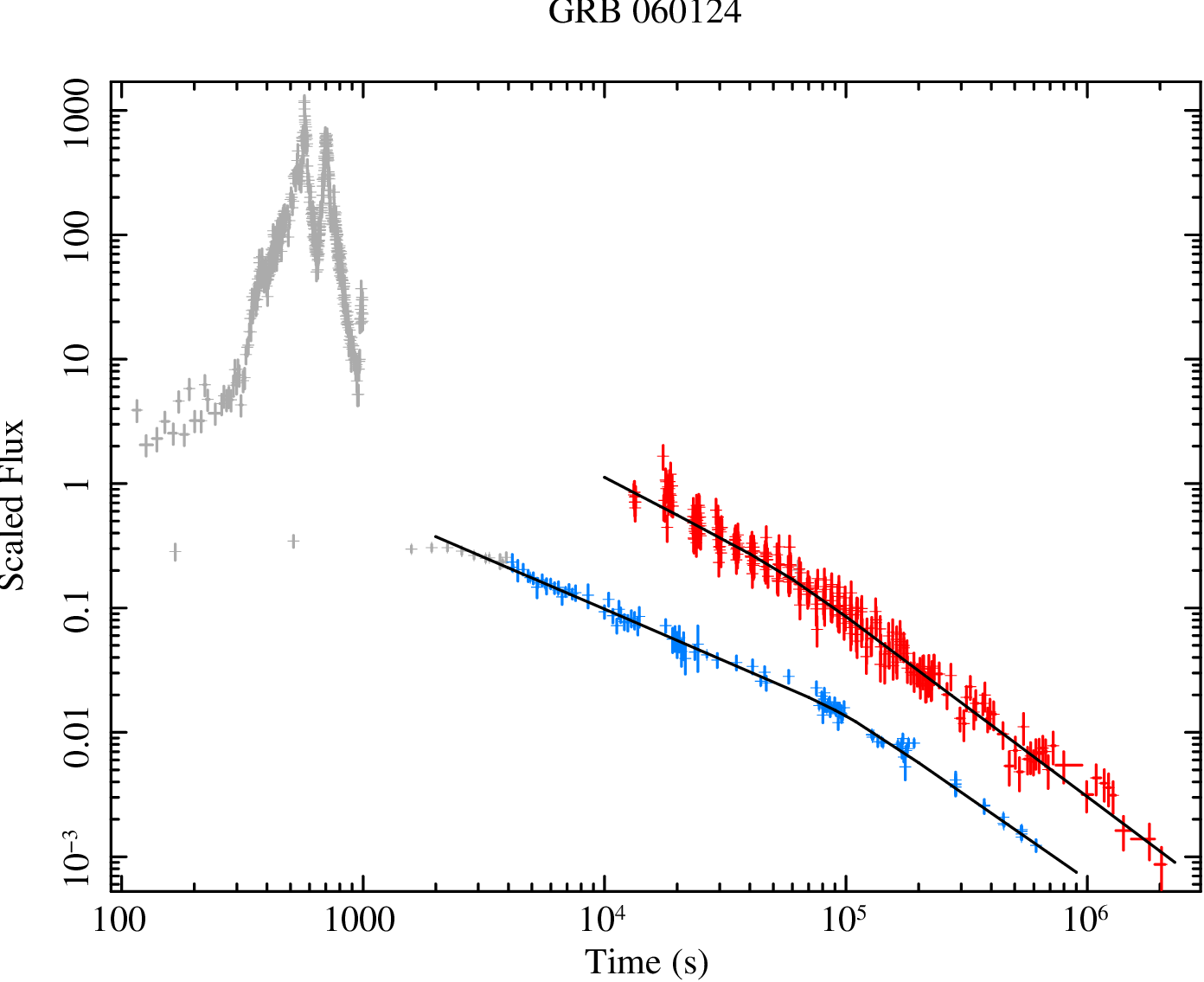} }
\end{minipage}
\caption{The power-law fits to the XRT (upper data) and optical (lower data) light curves of each of the bursts in our sample. Data in grey were not used in the power-law fits. Optical hosts, where available, were included in the models.} 
\label{breaks:lightcurves1} 
\end{figure*}

\setcounter{figure}{0}
\begin{figure*}
 \centering 
\begin{minipage}{80mm}
\resizebox{80mm}{!}{\includegraphics[angle=-90]{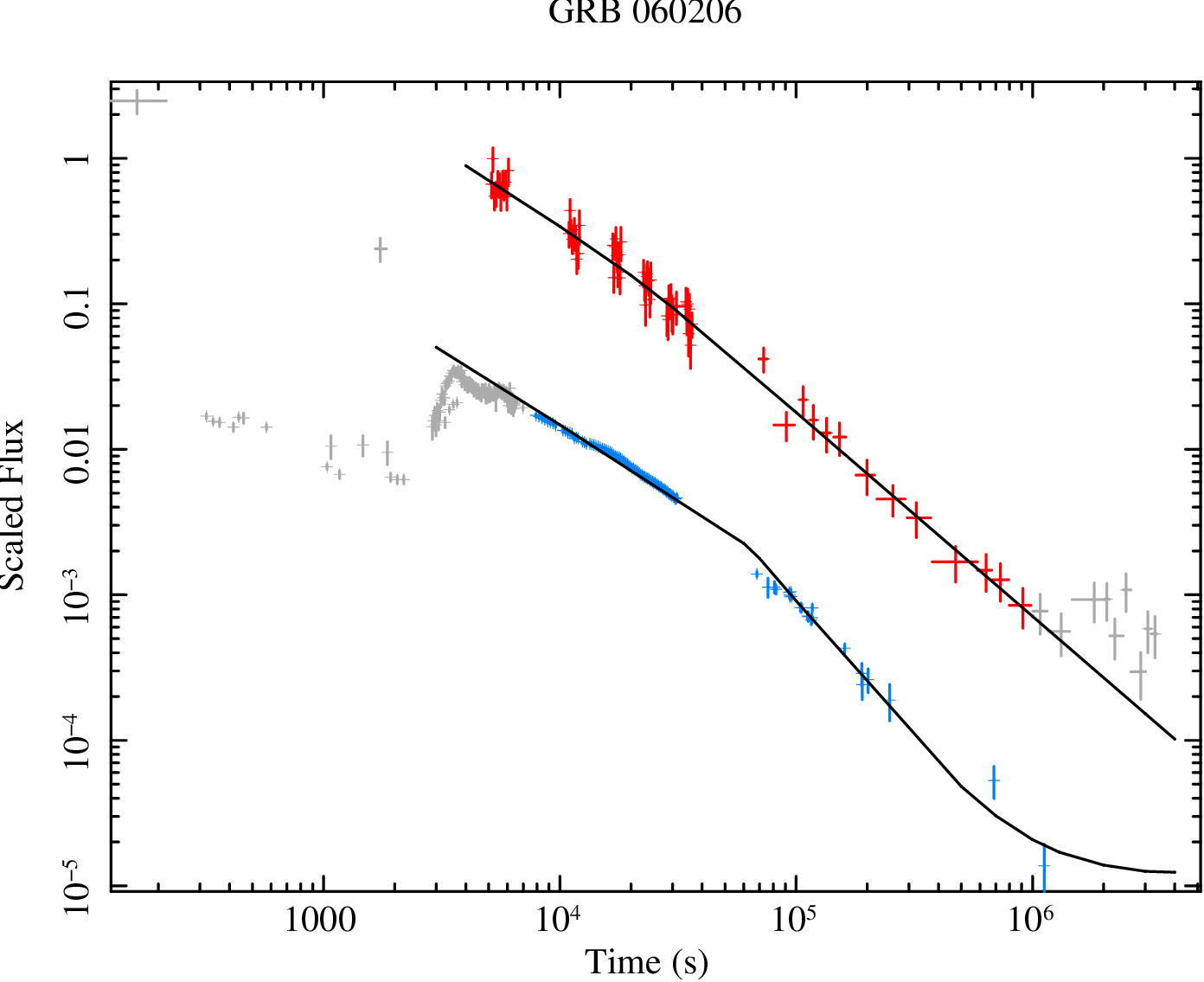} }
\end{minipage}
\hspace{3mm}
\vspace{3mm}
\begin{minipage}{80mm}
\resizebox{80mm}{!}{\includegraphics[angle=-90]{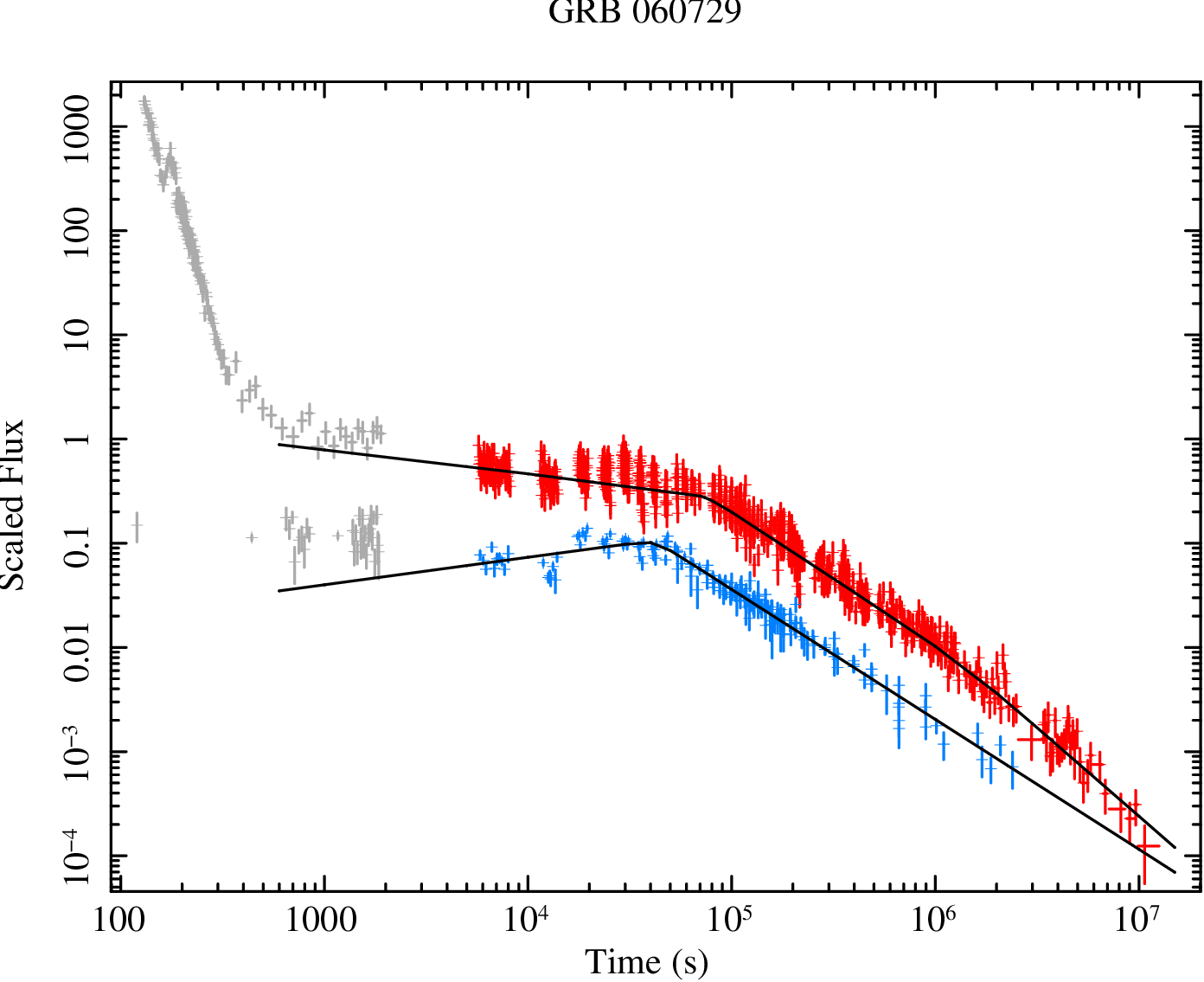} }
\end{minipage}
\vspace{3mm}
\begin{minipage}{80mm}
\resizebox{80mm}{!}{\includegraphics[angle=-90]{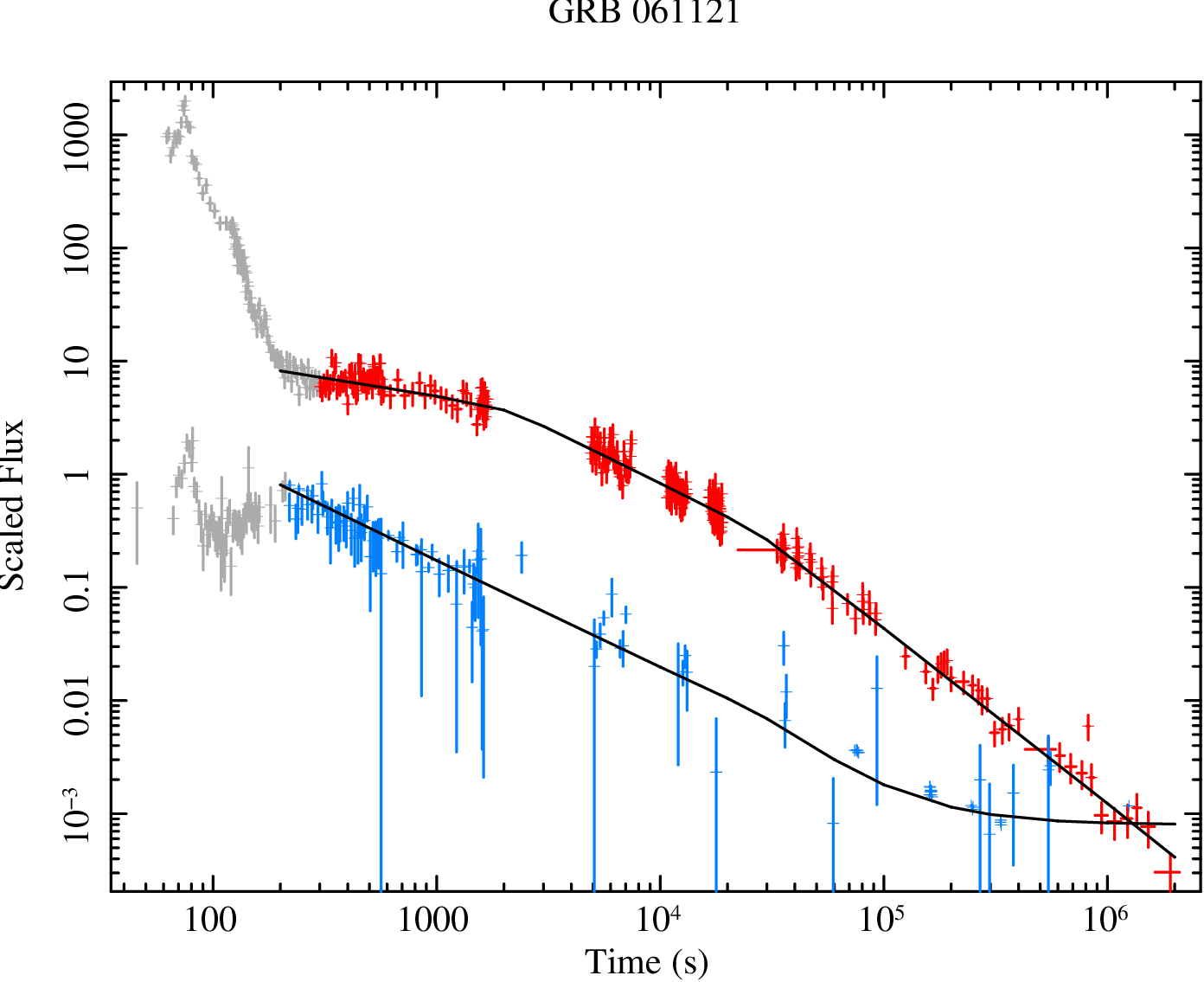} }
\end{minipage}
\hspace{3mm}
\begin{minipage}{80mm}
\resizebox{80mm}{!}{\includegraphics[angle=-90]{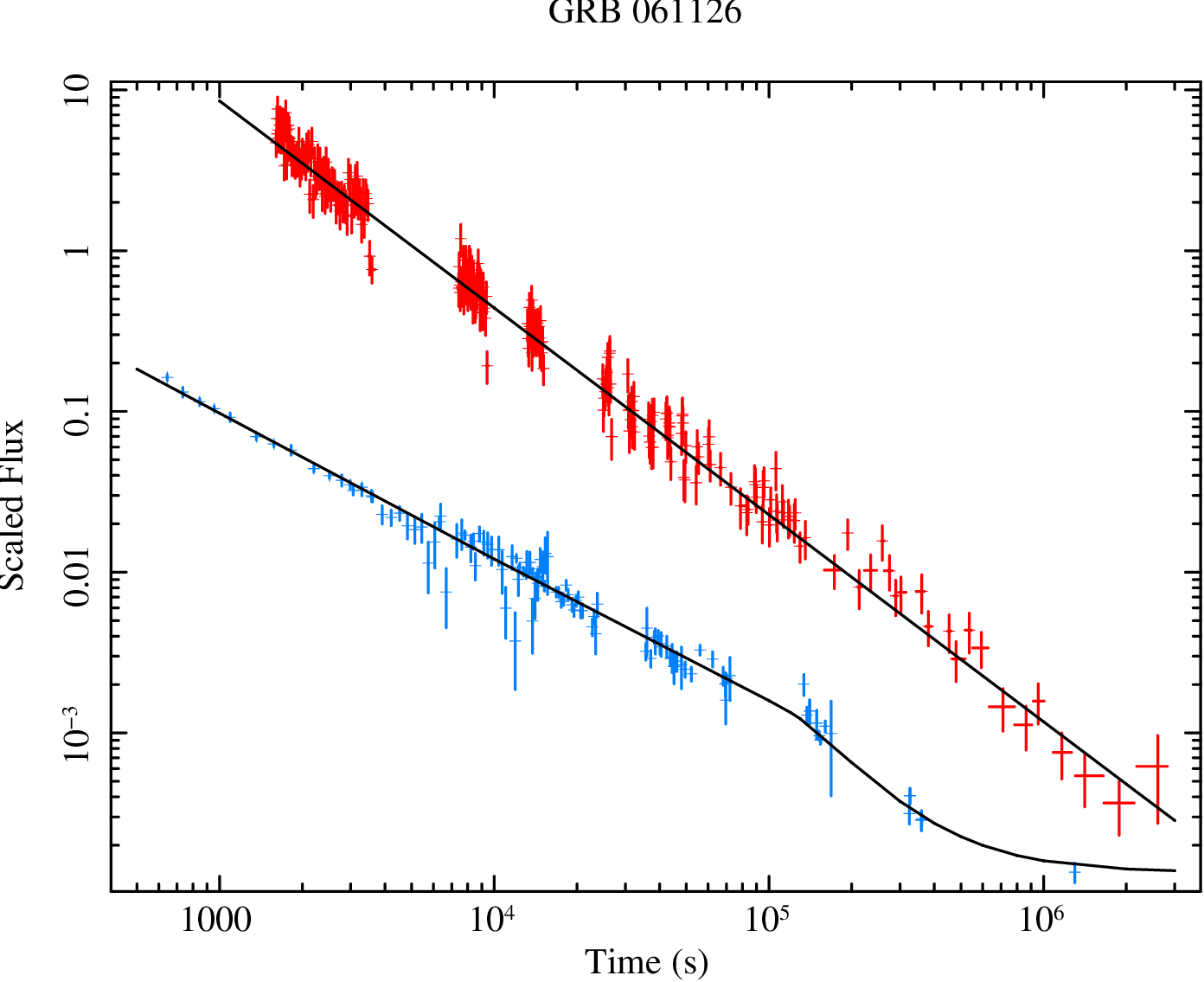} }
\end{minipage}
\caption{Continued.}
\label{breaks:lightcurves2} 
\end{figure*}

\begin{table}	
  \centering	
  \caption{Galactic absorption, $N_{{\rm H}}$ \citep{kalberla2005:A&A440}, Galactic extinction, $E_{(B-V)}$ \citep{schlegel1998:ApJ500}, and spectroscopic redshift, $z$, for the GRBs in our sample.} 	
  \label{breaks:sample-table} 	
  \begin{tabular}{l l l l } 
    \hline 
    GRB & $N_{{\rm H}}$	     & $E_{(B-V)}$  &   $z$\\ 
        & $\times 10^{20}$\,cm$^{-2}$ &              & \\
    \hline 
    050730	& 2.99	& 0.051	& 3.967$^{a}$        \\ 
    050801	& 6.33	& 0.096	& 1.56$^{\dagger}$$^{b}$ \\ 
    050802	& 1.86	& 0.021	& 1.71$^{c}$         \\ 
    050820A	& 4.41	& 0.044	& 2.6147$^{d}$       \\ 
    051109A	& 16.1	& 0.190	& 2.346$^{e}$        \\ 
    060124	& 8.98	& 0.135	& 2.297$^{f}$        \\ 
    060206	& 0.889	& 0.012	& 4.048$^{g}$        \\ 
    060729	& 4.49	& 0.054	& 0.54$^{h}$         \\ 
    061121	& 3.99	& 0.046	& 1.314$^{i}$         \\ 
    061126	& 10.2	& 0.182	& 1.1588$^{j}$        \\ 
    \hline 
  \end{tabular}
  \begin{list}{}{}
  \item[]$^{\dagger}$ photometric redshift
  \item[]
    $^{a}$\citet{chen2005:GCN3709}, 
    $^{b}$\citet{depasquale2007:MNRAS377}, 
    $^{c}$\citet{fynbo2005:GCN3749}, 
    $^{d}$\citet{ledoux2005:GCN3860},  
    $^{e}$\citet{cenko2006:GCN4592}, 
    $^{f}$\citet{quimby2005:GCN4221},  
    $^{g}$\citet{fynbo2006:A&A451}, 
    $^{h}$\citet{thoene2006:GCN5373},  
    $^{i}$\citet{bloom2006:GCN5826}, 
    $^{j}$\citet{perley2008:ApJ672}.
  \end{list}
\end{table}

\begin{table}	
  \centering	
  \caption{Optical references for the GRBs in our sample.} 	
  \label{breaks:optical_table} 	
  \begin{tabular}{l l l } 
    \hline 
    GRB & Filters & Reference\\ 
    \hline 
    050730	& $B,V,R,r^{\prime},I,i^{\prime},J,K$       &  1\\ 
    050801	& $UVM2,UVW1,U,V,R_{{\rm C}}$              &  2 \\ 
    050802	& $UVW2,UVM2,UVW1,U,B,V $                     &  3\\ 
    050820A	& $U,B,V,g,R,I_{C},z^{\prime}$     &  4 \\ 
    051109A	& $r_{R},C_{R}$	  &   5 \\ 
    060124	& $B,V,R_{C},I_{C}$          &  6 \\ 
    060206	& $V,R,I,J,H,K_{S} $              &  7 \\ 
    060729	& $UVW2,UVM2,UVW1,U,B,V	$                  &  8 \\ 
    061121	& $UVW2,UVM2,UVW1,U,B,V,R$	           &  9 \\ 
    061126	& $B,R_{C},r^{\prime},i^{\prime}$    &  10 \\ 
    \hline 
  \end{tabular}
  \begin{list}{}{}
  \item[]
    1 \citet{pandey2006:A&A460},
    2 \citet{rykoff2006:ApJ638,depasquale2007:MNRAS377},
    3 \citet{oates2007:MNRAS380},
    4 \citet{cenko2006:ApJ652},
    5 \citet{yost2007:ApJ657},
    6 \citet{curran2006:astro.ph.10067,misra2007:A&A464}; Kann et al. (in preparation),
    7 \citet{curran2007:MNRAS381,alatalo2006:GCN4702,stanek2007:ApJ654,wozniak2006:ApJ642},  
    8 \citet{grupe2007:ApJ662},
    9 \citet{page2007:ApJ663,halpern2006:GCN5853,halpern2006:GCN5851,halpern2006:GCN5847,halpern2006:GCN5840,melandri2006:GCN5827,yost2006:GCN5824},
    10 \citet{gomboc2008:ApJ687}.
  \end{list}
\end{table}

\subsection{X-ray}\label{breaks:xrt-obs}

The XRT event data for our sample bursts were initially processed with the FTOOL, \texttt{xrtpipeline (v0.11.4)}. Bad columns and pile-up were tested for and corrected for where necessary. Source and background spectra from the Photon Counting and Windowed Timing mode data (PC and WT; \citealt{hill2004:SPIE5165}) were extracted using suitable extraction regions. The PC mode source extraction regions were circular (30 pixel radius), circular with a central exclusion radius for sequences suffering pile-up, or circular with exclusion radii to exclude nearby sources which would have contaminated the spectra. WT mode source extraction regions were rectangular, and WT mode data did not suffer from pile-up during the time ranges for which we extracted spectra. Spectra were binned to have $\geq$20 photons per bin and the \texttt{v010} response matrices were used. 

Our light curve analyses are carried out on the pre-reduced, XRT light curves from the on-line repository \citep{evans2007:A&A469}. These light curves were fit with single or broken power-laws \citep{beuermann1999:A&A352} where the early, flaring data were eliminated from the fit. For bursts where there was a possible light curve break (Figure\,\ref{breaks:lightcurves1}), spectra were extracted pre-break (early) and post-break (late). The log mid-time of these spectra, $t_{{\rm early}}$ and $t_{{\rm late}}$, are given in Table\,\ref{breaks:sed-table}. 
If there was no possible break, early and late spectra were extracted for times before and after the approximate log mid-time of the afterglow, to test for any possible spectral change.
The early and late spectra (0.3-10.0\,keV) were fit both  individually and simultaneously (early+late) with absorbed power-laws in \texttt{Xspec 11.3.2} (Table \ref{breaks:Xspec_table}). $\chi^{2}$ statistics were used and the Galactic values of Column Density, $N_{{\rm H}}$, were fixed to the values of \citet{kalberla2005:A&A440} (Table \ref{breaks:sample-table}).

\subsection{Optical}

Optical photometric points in various bands were taken from the literature (Table \ref{breaks:optical_table}) and combined via a simultaneous temporal fit as detailed in \citet{curran2007:A&A467}. This fitting allowed us to find the common temporal slope of the optical data and the colour differences between bands. Using these colours, the optical data were then shifted to a common magnitude and converted into an arbitrary, scaled flux to produce joint optical light curves (Figure\,\ref{breaks:lightcurves1}). These light curves were fit with single or broken power-laws \citep{beuermann1999:A&A352}, including host galaxy contributions where known (Table\,\ref{breaks:temporal_indices}). Early times at which the underlying behaviour was ambiguous, or flaring, were excluded from the fit. Where we deemed the errors to be unrealistically small and where they gave unacceptable values of $\chi^{2} >> 1$, they were increased so as to give  $\chi^{2} \sim 1$, specifically in GRBs 050730 ($\times 5$), 050801 ($\times 2$), and 050820A ($\times 2$).

\subsection{Broadband SEDs} \label{breaks:sed-section}

The fit of optical colour differences and X-ray spectra described above were used to produce broadband spectral energy distributions (SEDs) at early and late times for each burst in our sample. Optical observations were interpolated to the log mid-time of the X-ray spectra ($t_{{\rm early}}$, $t_{{\rm late}}$), i.e., only bands with observations closest to that time were included in the SED. The optical SED points are corrected for Galactic extinction, $E_{(B-V)}$ \citep{schlegel1998:ApJ500}, if not already done so, and for line blanketing associated with the Lyman forest \citep{madau1995:ApJ441}. We used the simultaneous fitting method detailed in \cite{starling2007:ApJ661}, where the optical to X-ray SED is fit in count-space, having the advantage that no model for the X-ray data need be assumed a priori.  We fit the SEDs using models consisting of an absorbed single or broken power-law (SPL or BPL) with slopes tied to $\beta_1 = \beta_2 - 0.5$, as expected for a cooling break in the blast wave model \citep{meszaros1998:ApJ499,sari1998:ApJ497}.
We also use this data to calculate the spectral index between the optical at the lowest available frequency and the X-ray at 1.73\,keV, $\beta_{{\rm opt-X}}$ (Table\,\ref{breaks:Xspec_table}) to compare with the predicted values of an SPL, $\beta_{{\rm opt-X}} = \beta_{{\rm X}}$, or a  BPL, $\beta_{{\rm opt-X}} = \beta_{{\rm X}} - 0.5$.

We use this broadband information in conjunction with the previously published SED analyses to estimate whether or not there was a break between the two regimes (Table\,\ref{breaks:sed-table}). When the $\chi^2$ of the SPL and BPL fits to the broadband SED are approximately 1, or the optical spectral index is consistent with the X-ray spectral index, we are unable to say if the broken power-law is a significantly better fit, so consider both cases.


\section{Results}\label{breaks:results}

\begin{table*}	
  \centering	
  \caption{F-test probabilities, F$_{{\rm prob}}$, for the fits of the early time, $t_{{\rm early}}$, and late time, $t_{{\rm late}}$, broadband SEDs and whether they are best described by a single or a broken power-law (SPL or BPL).} 	
  \label{breaks:sed-table} 	
  \begin{tabular}{l l l l l l l} 
    \hline 
    GRB & $t_{{\rm early}}$ &  F$_{{\rm prob}}$ & & $t_{{\rm late}}$ &  F$_{{\rm prob}}$  & \\ 
    & s &  &  & s &   & \\ 
    \hline
    050730	& $7.03 \times 10^{3}$ & 0.38 & & $8.77 \times 10^{4}$ & 1.0   & SPL/BPL$^{\dagger}$  \\ 
    050801	& 161 & 0.20 & & & --  & SPL/--  \\ 
    050802	& $1.18 \times 10^{3}$ & 0.01  & &  $8.88 \times 10^{4}$ & $8 \times 10^{-4}$  & SPL/BPL    \\ 
    050820A	& $1.66 \times 10^{4}$ & $3 \times 10^{-14}$ & & $7.59 \times 10^{35}$ & 0.63  & --/BPL   \\ 
    051109A	& $8.96 \times 10^{3}$ & 1.0   & &  $2.80 \times 10^{5}$ & 1.0  & SPL/BPL$^{\dagger}$   \\                             
    060124	& $3.19 \times 10^{4}$ & $3 \times 10^{-21}$ & & $2.61 \times 10^{5}$ & $9 \times 10^{-9}$  & --/BPL   \\ 
    060206	& $1.00 \times 10^{4}$ & $1.9 \times 10^{-3}$ & & $8.20 \times 10^{4}$ & 0.63  & SPL/BPL   \\ 
    060729	& $2.09 \times 10^{4}$ & $9 \times 10^{-11}$ & & $1.03 \times 10^{6}$ & $3 \times 10^{-5}$  & --/BPL  \\ 
    061121	& $9.66 \times 10^{3}$ & $6 \times 10^{-7}$ & & $2.63 \times 10^{5}$	& 0.02  & --/BPL   \\ 
    061126	& $2.59 \times 10^{3}$ & $6 \times 10^{-7}$ & & $1.44 \times 10^{5}$	& 1.0  & --/BPL   \\ 
    \hline 
  \end{tabular}
  \begin{list}{}{}
  \item[]$^{\dagger}$ The $\chi^2$ of the SPL and BPL are approximately 1 so we cannot state if the improvement is significant, despite the F-test.
  \end{list}
\end{table*}

\begin{table*}	
  \centering	
  \caption{Temporal fits of the light curves in our sample; pre-break temporal index, break time in seconds and post-break index for the X-ray and the optical light curves.} 	
  \label{breaks:temporal_indices} 	
  \begin{tabular}{l l l l l l l l } 
    \hline 
    GRB & $\alpha_{{\rm X,1}}$ & $t_{{\rm X,break}}$ &  $\alpha_{{\rm X,2}}$  &  
    & $\alpha_{{\rm opt,1}}$ & $t_{{\rm opt,break}}$ &  $\alpha_{{\rm opt,2}}$ \\ 
       &  & s& &  & & s&    \\ 
    \hline 
    050730	& 0.91 $\pm$ 0.07 & $(1.06 \pm 0.03) \times 10^{4}$ & 2.56 $\pm$ 0.04 & 
    &0.55 $\pm$ 0.03 &  1.4$^{+1.2}_{-0.3}$ $\times 10^{4}$ & 1.70$^{+0.4}_{-0.12}$   \\ 
    050801	& -0.6 $\pm$ 0.4  & 290$^{+50}_{-30}$  &   1.39 $\pm$ 0.04 & 
    & 0.13 $\pm$ 0.03 & 230 $\pm$ 10 & 1.16 $\pm$ 0.01  \\ 
    050802	& 0.70 $\pm$ 0.04 & $(7.8 \pm 0.1) \times 10^{3}$ & 1.58  $\pm$ 0.04 & 
    & 0.81 $\pm$ 0.04 & -- & --   \\ 
    050820A	& 1.17 $\pm$ 0.02 & $(5.9 \pm 1.6) \times 10^{5}$ & 1.72 $\pm$ 0.15  & 
    & 0.91 $\pm$ 0.01 & $(5.5 \pm 1.0) \times 10^{5}$ & 1.60 $\pm$ 0.12   \\ 
    051109A	& 1.07 $\pm$ 0.03 & $(8 \pm 4) \times 10^{4}$ & 1.36 $\pm$ 0.05 & 
    &  0.66 $\pm$ 0.01 & $(5.7 \pm 0.7) \times 10^{4}$ & 1.54 $\pm$ 0.13  \\ 
    060124	&  1.01 $\pm$ 0.08 & $(5.8 \pm 1.0) \times 10^{4}$  & 1.45 $\pm$ 0.03  & 
    & 0.83 $\pm$ 0.01 & $1.0^{+0.1}_{-0.07}\times 10^{5}$ & 1.35 $\pm$ 0.02  \\ 
    060206	&  1.04 $\pm$ 0.10 & $2.2^{+2.0}_{-0.8} \times 10^{4}$ & 1.40 $\pm$ 0.07 & 
    & 1.022 $\pm$ 0.006 & $(6.23 \pm 0.05) \times 10^{4}$ & 1.99 $\pm$ 0.05  \\ 
    060729$^{a}$ & 0.23 $\pm$ 0.02 &  $(7.4 \pm 0.3) \times 10^{4}$ & 1.28 $\pm$ 0.02 & 
    & -0.26 $\pm$ 0.02  & $(4.2 \pm 0.1) \times 10^{4}$ & 1.28 $\pm$ 0.02  \\ 
    061121$^{b}$ & 0.97 $\pm$ 0.02 & $(2.9 \pm 0.4) \times 10^{4}$ & 1.55 $\pm$ 0.03 & 
    & 0.96 $\pm$ 0.02 & 2.9 $\times 10^{4}$ (fixed)  & 1.55 (fixed)  \\ 
    061126	& 1.29 $\pm$ 0.01 & -- & -- & 
    &  0.91 $\pm$ 0.01 & $(1.33 \pm 0.08) \times 10^{5}$  & 1.9 $\pm$ 0.1  \\ 
    \hline 
  \end{tabular}
  \begin{list}{}{}
  \item[]$^{a}$ Fitting a doubly broken power-law to the X-ray, as plotted, also reveals a late break at $(1.4 \pm 0.4) \times 10^{6}$\,s to $\alpha_{{\rm X,3}} = 1.70 \pm 0.09$.
  \item[]$^{b}$ Fitting a doubly broken power-law to the X-ray, as plotted, also reveals an early break at $2000 \pm 400$\,s from $\alpha_{{\rm X,0}} = 0.32 \pm 0.04$.
  \end{list}
\end{table*}

\begin{table*}	
  \centering	
  \caption{X-ray spectral fits for the GRBs in our sample; spectral indices, $\beta_{{\rm X}}$, and host extinctions, $N_{{\rm H}}$, for early and late time (as defined in Table\,\ref{breaks:sed-table}) X-ray spectra (see section \ref{breaks:xrt-obs}). The difference between $\beta_{{\rm X,early}}$ and  $\beta_{{\rm X, late}}$, $\Delta\beta_{{\rm X}}$, is given as the number of $\sigma$, where $\sigma$ is the quadrature sum of the errors.The optical to X-ray spectral index, $\beta_{{\rm opt-X}}$, is also given.} 	
  \label{breaks:Xspec_table} 	
  \begin{tabular}{l l l l l l l l l l l} 
    \hline 
    GRB & $\beta_{{\rm X, early}}$ & $N_{{\rm H, early}}$ & &  $\beta_{{\rm X, late}}$  & $N_{{\rm H, late}}$ & & $\Delta\beta_{{\rm X}}$  & $\beta_{{\rm X, early+late}}$  & $N_{{\rm H, early+late}}$  & $\beta_{{\rm opt-X}}$ \\ 
        &   &  $\times 10^{22}$\,cm$^{-2}$ & &   & $\times 10^{22}$\,cm$^{-2}$ & & $\sigma$ & & $\times 10^{22}$\,cm$^{-2}$  &   \\
    \hline 
    050730	& 0.56 $\pm$ 0.03 & 1.1 $\pm$ 0.2   && 0.73 $\pm$ 0.03 & 0.90 $\pm$ 0.3  && 4.0  &  0.63 $\pm$ 0.02 & 1.08 $\pm$ 0.17 & 0.85$\pm$0.06   \\ 
    050801	&  0.9 $\pm$ 0.2  &    $\leq$ 1.0   && 0.79 $\pm$ 0.10 &     $\leq$ 4.25 && 0.5  &  0.82 $\pm$ 0.10 & $\leq$ 0.24     & 1.08$\pm$0.16 \\ 
    050802	& 0.75 $\pm$ 0.04 & 0.28 $\pm$ 0.06 && 0.89 $\pm$ 0.04 & 0.30 $\pm$ 0.07 && 2.5  &  0.81 $\pm$ 0.03 & 0.29 $\pm$ 0.05 & 0.61$\pm$0.10   \\ 
    050820A	& 1.03 $\pm$ 0.04 & 0.80 $\pm$ 0.12 && 0.90 $\pm$ 0.05 & 0.29 $\pm$ 0.17 && 2.0  &  1.00 $\pm$ 0.03 & 0.68 $\pm$ 0.10 & 0.37$\pm$0.15  \\ 
    051109A	& 1.04 $\pm$ 0.04 & 1.1  $\pm$ 0.2  && 1.05 $\pm$ 0.09 & 1.0 $\pm$  0.45 && 0.1  &  1.04 $\pm$ 0.04 & 1.1 $\pm$ 0.2   & 0.73$\pm$0.13 \\ 
    060124	& 1.02 $\pm$ 0.04 & 0.91 $\pm$ 0.12 && 1.00 $\pm$ 0.04 & 0.90 $\pm$ 0.13 && 0.4  &  1.01 $\pm$ 0.03 & 0.91 $\pm$ 0.10 & 0.74$\pm$0.10  \\ 
    060206	& 1.26 $\pm$ 0.08 & 1.2 $\pm$ 0.4   && 0.92 $\pm$ 0.13 & 1.3  $\pm$ 0.7  && 2.2  &  1.17 $\pm$ 0.08 & 1.2 $\pm$  0.4  & 0.90$\pm$0.08 \\ 
    060729	& 1.11 $\pm$ 0.03 & 0.16 $\pm$ 0.01 && 1.12 $\pm$ 0.02 & 0.25 $\pm$ 0.01 && 0.3  &  1.11 $\pm$ 0.02 & 0.20 $\pm$ 0.1  & 0.92$\pm$0.08 \\ 
    061121$^{a}$& 1.01 $\pm$ 0.05 & 0.79 $\pm$ 0.09 && 0.83 $\pm$ 0.04 & 0.68 $\pm$ 0.10 && 2.8  &  0.94 $\pm$ 0.04 & 0.74 $\pm$ 0.08 & 0.59$\pm$0.21 \\ 
    061126	& 0.84 $\pm$ 0.06 & 0.68 $\pm$ 0.13 && 0.95 $\pm$ 0.03 & 0.69 $\pm$ 0.08 && 1.6  &  0.91 $\pm$ 0.04 & 0.69 $\pm$ 0.08 & 0.50$\pm$0.29 \\ 
    \hline 
  \end{tabular}
  \begin{list}{}{}
  \item[]$^{a}$ For the first decay segment, before 2000\,s we find $\beta_{{\rm X}} = 0.99 \pm 0.07$ and $N_{{\rm H}} = (0.78 \pm 0.13) \times 10^{22}$\,cm$^{-2}$. 
  \end{list}
\end{table*}

Our results of the temporal (Figure\,\ref{breaks:lightcurves1}) and X-ray spectral fits are summarised in Tables \ref{breaks:temporal_indices} and \ref{breaks:Xspec_table}, and the details for individual bursts are discussed below. We defer the interpretation of these results to section \ref{breaks:discussion}.

\subsection{Individual bursts}

\subsubsection{GRB\,050730}

There is a clear break in both the optical and X-ray light curves at approximately the same time ($\sim 1 \times 10^4$\,s). Fits of the X-ray light curve before this break are complicated by the lack of data and the possibility that the early flaring may continue to the break, hence the pre-break X-ray decay of $\alpha_{{\rm X,1}} = 0.91 \pm 0.07$ should be treated with caution and as an upper limit. 
Similar deviations from power-law decay are observed in the optical but a greater time range of data before the break implies that the underlying slope here is less likely to deviate from the measured value. After the break the lack of optical data is reflected in the fit errors of the slope, which may be steeper than measured if there is a significant flux contribution from a host (for which there is no estimate). 

There is an apparent X-ray spectral change across the break but the sparsity of optical data after the break makes this difficult to confirm. From our  broadband fits we cannot make a conclusive statement on the existence of a spectral break but a single power-law offers an reasonable fit in agreement with our consistent X-ray and optical to X-ray spectral indices. \citet{pandey2006:A&A460} find no evidence for a spectral break between the X-ray and optical, with an optical spectral index equal to that of the X-ray, at least before the break.

\subsubsection{GRB\,050801}

A simultaneous break is observed in both bands and the optical light curve is well constrained both before and after the break. There is no detectable late break in the optical, as the last data point is consistent with the extrapolation from early times. The X-ray light curve is less well constrained due to few data points. Fitting the decaying X-ray data, after its peak at 290\,s, with a single power-law, as opposed to a broken power-law including the apparent rising segment, gives a significantly shallower decay of $\alpha_{{\rm X,2}} = 1.16 \pm 0.03$, in agreement with the optical.
There is no measurable change in the X-ray spectral index over the break. Our SED fits are consistent with a SPL between the spectral regimes, in agreement with our optical to X-ray spectral indices being consistent with the X-ray spectral index. Both \citet{rykoff2006:ApJ638} and \citet{depasquale2007:MNRAS377} also find an X-ray to optical spectral index consistent with that of the X-ray, implying that there is no spectral break between the two regimes.

\subsubsection{GRB\,050802}

In agreement with the results found by \citet{oates2007:MNRAS380}, the X-ray light curve is well described by a broken power-law of $\alpha_{{\rm X,1}} = 0.70 \pm 0.04$ and $\alpha_{{\rm X,2}} = 1.58 \pm 0.04$, while the optical is described by a single power-law of $\alpha_{{\rm opt,1}} = 0.81 \pm 0.04$. If a broken power-law is fit to the optical, the fit returns $\alpha_{{\rm opt,1}} = \alpha_{{\rm opt,2}}$, however the slopes and break time of the X-ray light curve may be forced, to increase the  $\chi^2$ from 0.89 to 1.47. This may be an acceptable fit since we cannot apply a correction for a host galaxy as it was not observed.

We find no measurable change in the X-ray spectral index over the temporal break. From our broadband SED fits  it is not possible to rule out a spectral break, though neither is one required. This is in agreement with  an X-ray to optical spectral index consistent with the X-ray spectral index as both we and  \citet{oates2007:MNRAS380} find.

\subsubsection{GRB\,050820A}

After an initial optical rise and X-ray flaring, both light curves are well fit by broken power-laws with consistent break times and post break slopes. 
According to \citet{cenko2006:ApJ652} a host contribution may affect the last optical (HST) points at $\sim 3 \times 10^6$\,s so the fit decay and break time should possibly be treated as  lower limits. We find no need to fit a decaying segment from 500 - 5000\,s, as they have done, as an extrapolation of the later data back to this time does not support a break. 

Our broadband SED fits and our derived X-ray and optical to X-ray spectral indices support a break between the two spectral regimes. This is in agreement with \citet{cenko2006:ApJ652} who find an optical spectral index, $\beta_{{\rm opt}} \sim 0.67 \pm 0.08$, $3.9\sigma$ shallower than the X-ray spectral index of $\beta_{{\rm X}} = 1.00 \pm 0.03$.

\subsubsection{GRB\,051109A}

There is a clear break in the optical light curve and even though it is not obvious, an F-test probability of $5 \times 10^{-5}$, supports a break at a consistent time in the X-rays, with a confidence of almost 5$\sigma$. Fixing the X-ray break time to the optical break time only affects the decay indices within uncertainties. 
From our broadband SED, as well as our  X-ray and optical to X-ray spectral indices, we find both single and broken power-laws offer acceptable fits. This is in agreement with  \citet{yost2007:ApJ657} who find an X-ray to optical spectral index of  $\sim 0.7 \pm 0.2$, only $\sim 1.5\sigma$ lower than the X-ray spectral index of $\beta_{{\rm X}} = 1.04 \pm 0.04$.

\subsubsection{GRB\,060124}

After the initial X-ray flaring, when the optical is not sampled well enough to be able to judge its behaviour, both afterglows are well described as broken power-laws with consistent break times and post-break slopes of $\alpha_{{\rm opt,1}} = 1.35 \pm 0.02$ and $\alpha_{{\rm X,1}} = 1.45 \pm 0.03$. We find no X-ray spectral change over this break. Our broadband SED conclusively requires a spectral break between the two regimes, consistent with the ambiguous difference between the X-ray and optical to X-ray spectral indices.

\subsubsection{GRB\,060206}

After initial rising in the optical and flaring behaviour in the X-ray light curves, both decay as power-laws. The optical afterglow is clearly a broken power-law, while the X-ray is well fit by a single power-law of $\alpha = 1.28 \pm 0.02$. However, the X-ray can also be fit by a broken power-law if the data past $1 \times 10^{6}$\,s are eliminated. This is valid as there is significant contamination from a nearby source as discussed by \citet{curran2007:MNRAS381} and shown in the reduction of \citet{butler2007:ApJ663}. The post-break X-ray temporal index should hence be treated as a lower limit.
Neither the difference between the X-ray and optical to X-ray spectral indices nor the SED are conclusive as to whether a break is required but given that the F-test probability corresponds to a less than $3\sigma$ certainty we favour here, as we have previously \citep{curran2007:MNRAS381}, a single power-law. Neither do we find any change in X-ray spectral index over the break greater than 3$\sigma$.

\subsubsection{GRB\,060729}

Both light curves follow parallel slopes after an almost simultaneous break at $\sim 6 \times 10^{4}$\,s which shows no X-ray spectral change. Before this break, from 5000\,s, the X-ray light curve is decaying with $\alpha_{{\rm X,1}} = 0.23 \pm 0.02$, though this is somewhat dependent on  the time one chooses to start the fit. The optical flux over this time is increasing with $\alpha_{{\rm opt,1}} = -0.26 \pm 0.02$, though if the earlier optical data are included, this is reduced to $\alpha_{{\rm opt,1}} = -0.05^{+0.04}_{-0.06}$, leaving the post-break slope unchanged, within uncertainties, and causing the break time to increase but not significantly. 
Fitting a doubly broken power-law to the X-ray light curve, we also find a second break in the X-ray light curve at $(1.4 \pm 0.4) \times 10^{6}$\,s to $\alpha_{{\rm X,3}} = 1.70 \pm 0.09$. An F-test comparison of the $\chi^{2}$ of the doubly broken power-law and the broken power-law give a highly significant probability of $2.5 \times 10^{-7}$, which is much greater than 5$\sigma$. The optical light curve does not sample this time so is unable to confirm or refute the break, which was not identified in the light curve analysis of \citet{grupe2007:ApJ662}.

Our SED analysis finds a BPL is favoured over a SPL and hence there is likely a spectral break between the two regimes; in direct contrast to the parallel nature of the optical and X-ray light curves. However the early time SED, where the spectral break is most pronounced, is taken before the temporal break at $\sim 6 \times 10^{4}$\,s, where the light curves are not parallel; the late time SED, during the parallel phase, supports a BPL with less certainty. \citet{grupe2007:ApJ662} find a constant optical to X-ray spectral index of $\sim 0.8 \pm 0.05$, compared to our  X-ray spectral index of $\beta_{{\rm X}} = 1.11 \pm 0.02$ and optical to X-ray spectral index of  $\beta_{{\rm opt-X}} = 0.92 \pm 0.08$, all of which are consistent with a SPL. Given this we cannot conclusively support the BPL nature if the SED.

\subsubsection{GRB\,061121}

The X-ray light curve displays a clear steepening from early to late times which is modelled by a doubly broken power-law. This a significantly better fit than a broken power-law with a break at $\sim 9 \times 10^3$\,s (F-test probability $\sim 10^{-20}$). After a rise to 212\,s, the optical light curve appears to decay with a single power-law of $\alpha_{{\rm opt,1}} = 0.92 \pm 0.021$. However, when the effects of the host galaxy, which \citet{page2007:ApJ663} estimate to be $\sim 2.5\mu$Jy in the $R$ band, are taken into account, a break is detectable. There is not enough data after the break to constrain the optical break time or temporal decay so they are set at the X-ray values, which gives an acceptable fit. 

As the light curve displays 3 power-law segments, X-ray spectra were extracted for each (Table\,\ref{breaks:Xspec_table}). The first two segments are in excellent agreement with each other and in marginal agreement ($2.8 \sigma$) with the last segment. The first segment, and hence the second also, may be affected by flux from the early prompt emission. 
We find a significant improvement in the description of the optical to X-ray SED, when a broken power-law is used in favour of a single power law, in agreement with the optical to X-ray spectral index as well as the SED of \citet{page2007:ApJ663}.

\subsubsection{GRB\,061126}

Before the effect of the host is taken into account, the optical light curve is well described by a single power law of $\alpha_{{\rm opt}} = 0.930 \pm 0.005$. When the host is modelled, a break from  $\alpha_{{\rm opt,1}} = 0.91 \pm 0.01$ to  $\alpha_{{\rm opt,2}} = 1.9 \pm 0.1$ becomes clear. The X-ray light curve does not seem to be consistent with a break since when a broken power-law, with a break time equal to that of the optical, is fit we find $\alpha_{{\rm X,1}}  = \alpha_{{\rm X,2}}$. However, forcing the post-break slope to equal that of the optical only increases the $\chi^2_{\nu}$ from 1.35 to 1.60 and while there is no obvious source of contamination to the X-ray light curve, we cannot rule out a break.
We find in our broadband SED fit, as do \citet{gomboc2008:ApJ687}, that the optical to X-ray SED is best fit by a broken power-law at least before the break. The SED after the break and the optical to X-ray spectral index are consistent with this, though inconclusive.


\begin{table*}	
  \centering	
  \caption{A summary of the relationships for temporal and spectral indices, $\alpha$ and $\beta$, for a given value of electron energy distribution index, $p$, in the different regimes (from e.g. \protect\citealt{starling2008:ApJ672}, assuming $p>2$). The shallowing effect, $\Delta\alpha$, of continued energy injection from \citet{nousek2006:ApJ642}.}
  \label{breaks:p-relations}	
  \begin{tabular}{l c c  c} 
    \hline 
    &  $\alpha$ & $\Delta\alpha$ & $\beta$ \\
    \hline 
    $\nu_{{\rm c,m}}  < \nu$  ($k$)                   &  $(3p-2)/4$ & $q(2+p)/4$ & $p/2$      \\
    $\nu_{{\rm c,m}}  < \nu$  (jet)                   &   $p$       & -- & $p/2$      \\
    $\nu_{{\rm m}} < \nu < \nu_{{\rm c}}$ ($k$)     & $\frac{12(p-1)-k(3p-5)}{4(4-k)}$ & $q(3-k+p)/4$  &  $(p-1)/2$  \\
    $\nu_{{\rm m}} < \nu < \nu_{{\rm c}}$ ($k=0$)   &  $3(p-1)/4$ & $q(3+p)/4$ & $(p-1)/2$  \\
    $\nu_{{\rm m}} < \nu < \nu_{{\rm c}}$ ($k=2$)   &  $(3p-1)/4$ & $q(1+p)/4$ & $(p-1)/2$  \\
    $\nu_{{\rm m}} < \nu < \nu_{{\rm c}}$ (jet)     &  $p$        & -- & $(p-1)/2$  \\
    \hline  
  \end{tabular}
\end{table*}

\section{Discussion}\label{breaks:discussion}

We use the blast wave model \citep{rees1992:MNRAS258,meszaros1998:ApJ499} to describe the temporal and spectral properties of the GRB afterglow emission. We assume on-axis viewing, a standard jet structure and no evolution of the microphysical parameters. The relations between the temporal and spectral indices and the blast wave parameters that we use are summarised in Table \ref{breaks:p-relations} (see also e.g.  \citealt{starling2008:ApJ672} and \citealt{nousek2006:ApJ642}). 
Our general method was to estimate the value of the electron energy distribution index, $p$,  from the X-ray spectral index (section \ref{breaks:p-derivation}) and use this to calculate the predicted values of temporal decay (which  are sub-scripted with $p$, section \ref{breaks:interpret}). 
We derive $p$ from the spectral index as opposed to the temporal index since for a given spectral index there are only two possible values of $p$, while for a given temporal index there are multiple possible values. Spectral slopes are dependent only on $p$ and the position of the cooling break. Temporal indices, $\alpha$,  are dependent on  $p$, the position of the cooling break, the circumburst density profile, $k$, and on possible continued energy injection (Table\,\ref{breaks:p-relations}). We assume that continued energy injection takes the form  of $E \propto t^{q}$ where $q$ is dependent on the shallowing effect of the injection, $\Delta\alpha = \alpha_{p} - \alpha_{{\rm observed}}$, and on the value of $p$ (\citealt{nousek2006:ApJ642}; though they define $q$ as $E \propto t^{q+1}$). This injection period from $t_i$ to $t_f$ increases the energy of the afterglow by a factor of $f = (t_{f}/t_{i})^{q}$.
Temporal indices are also prone to being incorrectly estimated from broken power-law fits \citep{johannesson2006:ApJ640}, underestimating post-break indices.

\subsection{Derivation of $p$}\label{breaks:p-derivation}

The values of the electron energy distribution index, $p$, are derived from the X-ray spectral index of the afterglow,  $\beta_{{\rm X}}$ (Table \ref{breaks:Xspec_table}), following the relations in Table \ref{breaks:p-relations}.  As most of the sample show no evolution of X-ray spectral index, we used the average spectral index, $\beta_{{\rm early+late}}$ in all cases except one. For GRB\,050730 there was an apparent spectral evolution so we used only the late time spectral index, $\beta_{{\rm late}}$, since this was less likely to be contaminated by flux from the  prompt emission; in this case the late spectrum better overlaps with the time range where we apply the blast wave model to the light curve so it is more proper to use it alone.

For a given value of X-ray spectral index, there are two possible values of $p$ (Table \ref{breaks:p-table}) depending on whether the cooling break, $\nu_{{\rm c}}$, is below ($p = 2\beta$) or above ($p = 2\beta +1$) the X-ray frequencies, $\nu_{{\rm X}}$. If the optical to X-ray SED  does not display a break then the cooling break can either be above the X-ray regime or below the optical regime and the blast wave predictions of each $p$ are compared to the observed temporal slopes to discern which is correct. If the SED requires a broken power-law it most likely implies that a cooling break lies between the two regimes and is below the X-ray regime. A cooling break requires, or must be consistent with, a difference between the spectral slopes of $\Delta\beta = 0.5$. 
However, a break between the two regimes does not necessarily imply a cooling break; it may be due to the fact that each regime has a different spectral index since they are originating from different emission regions, as is claimed to be the case for GRB\,080319B \citep{racusin2008:Natur455}. 
A spectral break due to two different emission regions does not have a predictable difference between slopes but for this interpretation to work, one must be able to explain why each emission region is only visible in one regime. A cooling break is a more likely explanation in the majority of cases but a comparison of the blast wave predictions of each $p$ with the observed light curves is required.

\subsubsection{Broadband  versus X-ray}

We use the X-ray spectral index to derive the value of $p$ because the value of $\beta_{{\rm X}}$ is relatively unaffected by environmental unknowns. The main unknown in the X-ray spectral fits is that of the absorption due to the column density in the host galaxy; this is easily fit and only affects the low energies of the \emph{Swift} XRT spectra. X-ray spectra may also suffer contamination from nearby sources but this can be minimised through careful extraction of spectra. 
The optical spectral index is dependent on, and suffers somewhat of a degeneracy with, extinction in the host galaxy. 
The extinction affects the whole range of the optical wavelengths somewhat, while absorption only significantly affects less than about half a decade of the X-ray range. In fact, as the redshift increases, extinction has a greater effect on the optical bands, while absorption is shifted out of the X-ray range. 
The optical data ideally spans from the near infrared $K$ band to the ultraviolet $U$ band, offering approximately a decade of spectral range  with which to constrain the spectral index. This is a range similar to that offered by the XRT (0.3--10\,keV) but is rarely reached and, given the narrow range of  commonly observed bands ($UBV$ or $VRI$), may not be able to constrain the spectral index very well.

Broadband, optical to X-ray, spectra are used: to eliminate some of the difficulties associated with the optical data; increase the fitting range;  and reduce the number of free parameters by tying the optical and X-ray spectral indices to each other. 
This has proved successful in a number of cases and especially for \emph{BeppoSAX} ($\sim$0.1--10\,keV) bursts \citep{starling2008:ApJ672} where the quality of the X-ray spectra did not always allow the spectral index  to be sufficiently constrained without the optical data; 
the mean spectral index error of \emph{BeppoSAX} data  is 0.30 \citep{depasquale2006:A&A455}. 
\emph{Swift} XRT allows for more accurate X-ray spectral indices with the median error on XRT PC spectra being 0.11 \citep{evans2009:mnras} but this overestimates the error on many bursts of interest as \emph{Swift} detects many more faint bursts. In our sample the average spectral index error is only 0.043 or less than 5\% of the average spectral index of 0.94, from our early+late spectra.

While the spectral index should only be constrained further in a broadband fit, the difficulties associated with the combination of data from different telescopes, may introduce errors and incorrect estimates of $p$. 
There may also be an unknown or uncertain contribution from a host galaxy, or there may be undocumented errors on the zero-points used to convert from magnitude to flux. It is also worth noting that data used for the optical SEDs, if taken from literature, may have to be converted into different quantities for fitting and these conversions can add significant errors.
While broadband SEDs may offer excellent constraints on optical spectral indices, extinction and absorption, they are not generally necessary to constrain the X-ray spectral index and hence the possible values of $p$. They are however invaluable in the identification of spectral breaks between the two regimes which is necessary in the interpretation of a GRB within the blast wave model and the correct choice of $p$.

\begin{table}	
  \centering	
  \caption{Possible values of $p$, derived from the X-ray spectral index of the afterglow,  $\beta_{{\rm X}}$.} 	
  \label{breaks:p-table} 	
  \begin{tabular}{l l l l l} 
    \hline 
    GRB & $\nu_{{\rm c}} < \nu_{{\rm X}}$ & &  $\nu_{{\rm X}}< \nu_{{\rm c}}$  \\ 
    \hline
    050730	& 1.46 $\pm$ 0.06 & 	& 2.46 $\pm$ 0.06   \\ 
    050801	& 1.64 $\pm$ 0.20 & 	& 2.64 $\pm$ 0.20   \\ 
    050802	& 1.62 $\pm$ 0.06 & 	& 2.62 $\pm$ 0.06   \\ 
    050820A	& 2.00 $\pm$ 0.06 & 	& 3.00 $\pm$ 0.06   \\ 
    051109A	& 2.08 $\pm$ 0.08 & 	& 3.08 $\pm$ 0.08   \\ 
    060124	& 2.02 $\pm$ 0.06 & 	& 3.02 $\pm$ 0.06    \\ 
    060206	& 2.34 $\pm$ 0.16 & 	& 3.34 $\pm$ 0.16    \\ 
    060729	& 2.22 $\pm$ 0.04 & 	& 3.22 $\pm$ 0.04    \\ 
    061121	& 1.88 $\pm$ 0.08 & 	& 2.88 $\pm$ 0.08    \\ 
    061126	& 1.82 $\pm$ 0.08 & 	& 2.82 $\pm$ 0.08    \\ 
    \hline 
  \end{tabular}
\end{table}

\subsection{Blast wave interpretation}\label{breaks:interpret}

Here we compare the derived values of $p$ and the blast wave model predictions to the observed temporal and spectral properties to find the most likely parameters (Table\,\ref{breaks:blast-table}). Given the complexity and ambiguity of some of the temporal results summarised in Table \ref{breaks:temporal_indices}, we advise reference to the text of section \ref{breaks:results} also.

\subsubsection{GRB\,050730}

We find that $p=2.46 \pm 0.06$ for $\nu_{{\rm X}} < \nu_{{\rm c}}$ and unconstrained $k$ and $q$, though this in not unequivocal. 
The steep X-ray temporal decay of $\alpha_{{\rm X,2}} = 2.56 \pm 0.04$ in this burst can only be consistent with a post jet break decay of  $\alpha_{{\rm X,}p} = \alpha_{{\rm jet}} = p$ given $p = 2.46 \pm 0.06$ derived from the X-ray spectral index, assuming $\nu_{{\rm X}} < \nu_{{\rm c}}$. The optical slope while shallower is also consistent, within $2\sigma$, with a post jet break decay equal to $p$. 
This implies that parallel, pre-break slopes of between 1.10 and 1.60, depending on $k$, or shallower given energy injection are required for both regimes in disagreement with the $4\sigma$ difference between the observed slopes. Since we cannot confirm a possible steepening contribution to the X-ray light curve from flares, the interpretation of this burst and the derived value of $p$ is ambiguous.

\subsubsection{GRB\,050801}

We find that $p=2.64 \pm 0.20$ for $\nu_{{\rm X}} < \nu_{{\rm c}}$ and $k = -0.7 \pm 1.4$, $q = 0.65 \pm 0.15$.
If we adopt the value of post-break temporal X-ray slope from the single power-law, as opposed to that from the broken power-law, both regimes are parallel after the break with $\alpha_{\rm X} \sim \alpha_{\rm opt,1} = 1.16 \pm 0.01$, in agreement with the lack of a break in the SED. This implies  $\nu_{{\rm X}} < \nu_{{\rm c}}$ and $p = 2.64 \pm 0.20$, which predicts a slope of  $\alpha_{{\rm X,}p} = \alpha_{{\rm opt,}p} = 1.23 \pm 0.09$ for $k=0$, in agreement with the observed values. Calculating $k$ from the temporal slope and value of $p$ we get $k = -0.7 \pm 1.4$.  The break of $\Delta\alpha = 1.03 \pm 0.03$ observed in the optical, could then be due to energy injection at early times of $q = 0.65 \pm 0.15$, increasing the energy of the afterglow over th observed period of injection by a factor of $f \sim 5$. The behaviour of the X-ray at that time is unclear, though it does not rule out a break.

\subsubsection{GRB\,050802}

We find that $p=2.62 \pm 0.06$ for $\nu_{{\rm X}} < \nu_{{\rm c}}$ and $k = 1.7 \pm 0.3$, $q = 0.90 \pm 0.13$.
The post-break X-ray slope of $\alpha_{{\rm X,2}} = 1.58 \pm 0.04$ is consistent $p = 2.62 \pm 0.06$ ($\nu_{{\rm X}} < \nu_{{\rm c}}$) and a circumburst density profile of $k = 1.7 \pm 0.3$. Since it is not certain whether there is a break in the SED, the positioning of the X-ray frequency below the cooling frequency is valid. Given that we cannot rule out a temporal break in the optical light curve, it is also possible that $\alpha_{{\rm opt}} = \alpha_{{\rm X}}$ as required. The X-ray break of $\Delta\alpha = 0.88 \pm 0.06$, consistent with the optical, would then be due to energy injection  of $q = 0.9 \pm 0.13$ which increases the energy of the afterglow over the observed period of injection by a factor of $f \sim 10$.

If there is not a break in the optical, then a possible explanation may be that the two regimes are originating from different emission regions as proposed by \citet{oates2007:MNRAS380}, though this raises the question as to why only one regime is observed from each emission region.

\subsubsection{GRB\,050820A}

We do not find a confident blast wave interpretation consistent with the light curves of this burst. 
The inconsistency of the optical and X-ray slopes, in agreement with the broken power-law SED, suggests that the cooling frequency is between the two regimes. This would imply that $p = 2.00 \pm 0.06$ and $\alpha_{{\rm X,}p} = 1.00 \pm 0.03$, $4.7\sigma$ shallower than the observed pre-break value of $\alpha_{{\rm X,1}} = 1.17 \pm 0.02$. The optical, which is below the cooling frequency, is explained by a circumburst medium of density profile, $k = 1.0 \pm 0.3$. The post-break slopes are consistent with the predicted $\alpha_{{\rm jet}} = p = 2.00 \pm 0.06$ of this interpretation at the $3\sigma$ level and possibly better given that the optical may be affected by a unquantified host contribution. Given the $4.7\sigma$ difference between the predicted and observed X-ray slopes before the break we cannot confidently support this interpretation.

\subsubsection{GRB\,051109A}

We find that $p=2.08 \pm 0.08$ for $\nu_{{\rm c}} < \nu_{{\rm X}}$ and $k = -1.7 \pm 0.8$, $q=0$.
In this burst it is unclear whether or not there is a cooling break between the two regimes so we have considered both cases. If $\nu_{{\rm X}} > \nu_{{\rm c}}$ then $p = 2.08 \pm 0.08$ implies an X-ray temporal slope of $\alpha_{{\rm X,}p} = 1.06 \pm 0.04$, in agreement with the X-ray slope before the break. The optical is then below the cooling break and $p$ implies $\alpha_{{\rm opt,}p} = 0.81 \pm 0.03$ for $k=0$, which is greater than the observed value $\alpha_{{\rm opt}} = 0.66 \pm 0.01$ by $4.7\sigma$; in fact the observed temporal slope requires $k = -1.7 \pm 0.8$. The break, which is consistent in optical and X-rays is then interpreted as a jet break, which should go as $\alpha_{{\rm jet}} = p = 2.08 \pm 0.08$. The observed value of $\alpha \sim 1.4$ is shallower than expected but may be rolling over to the predicted value. 

Alternatively, if $\nu_{{\rm X}} < \nu_{{\rm c}}$ then $p = 3.08 \pm 0.08$ implies temporal slopes of $\alpha_{{\rm X,}p} = \alpha_{{\rm opt,}p} = 1.56 \pm 0.03$ with $k=0$, comparing well with the observed post-break slopes of both. The break can then interpreted as the cessation of continued energy injection. The difference between the observed pre-break indices and the values predicted by the blast wave are, on average, $\Delta \alpha \sim 0.7$, which corresponds to $q \sim 0.6$. 
However, the $13\sigma$ difference between these slopes, which should be parallel, is difficult to explain, making our first explanation more likely.

\subsubsection{GRB\,060124}

We find that $p=2.02 \pm 0.06$ for $\nu_{{\rm c}} < \nu_{{\rm X}}$ and $k = 0.0 \pm 0.4$, $q=0$.
The broken power-law SED and unequal pre-break decays require that the cooling break lies between the optical and X-ray frequencies, which implies $p = 2.02 \pm 0.06$. The predicted slopes of  $\alpha_{{\rm X,}p} = 1.02 \pm 0.03$ and $\alpha_{{\rm opt,}p} = 0.77 \pm 0.02$ ($k=0.0 \pm 0.4$) are in agreement with the observed values before the break. The observed post-break values of $\sim 1.4$ are shallower than the predicted value of $\alpha_{{\rm jet}} = p$, but may be rolling over to the asymptotic value, or may be shallower than expected due to a structured jet as suggested by Kann et al. (in preparation).

\subsubsection{GRB\,060206}

We find that either $p=2.34 \pm 0.16$ for $\nu_{{\rm c}} < \nu_{{\rm X}}$ with $q=0$ and $k$ unconstrained, or, $p=3.34 \pm 0.16$ for $\nu_{{\rm X}} < \nu_{{\rm c}}$ and $k = 1.3 \pm 0.7$, $q = 0.8 \pm 0.2$; the data are ambiguous.
The single power-law SED from optical to X-rays and the fact that both have equal temporal decay indices before the break, imply that both regimes are either above, or below, the cooling frequency.
In the first case $p = 2.34 \pm 0.16$, implying $\alpha_{p} = 1.26 \pm 0.09$ before the break, which is slightly steeper than, but within $3\sigma$ of the observed slopes which are $\alpha_{{\rm X,1}} = 1.04 \pm 0.10$ and $\alpha_{{\rm opt,1}} = 1.022 \pm 0.006$. The predicted post-break slope for a jet break, $\alpha_{{\rm jet}} = p$, could agree with the optical slope rolling over to an asymptotic value and with the `hidden' X-ray break. 

In the second case, the observed bands are below the cooling frequency and $p = 3.34 \pm 0.16$, implying $\alpha_{p}$ is between $1.76 \pm 0.08$ ($k=0$) and $2.26 \pm 0.11$ ($k=2$). The post-break optical slope, and post-hidden break X-ray slope, could then be caused by a circumburst density profile of $k = 1.3 \pm 0.7$. The pre-break slopes, which are shallower than predicted by $\Delta\alpha = 0.97 \pm 0.05$, could be caused by continued energy injection with $q = 0.8 \pm 0.2$, though this would increase the energy of the afterglow over the observed period of injection by a factor of $f \sim 30$. 
Given that we can equally well explain a cooling break either above or below the observed frequencies, we cannot say which of the two explanations is favoured.

\begin{table*}	
  \centering	
  \caption{Most likely values for the blast wave parameters after our interpretation of individual bursts (section \ref{breaks:interpret}). Where we interpret a jet break within the light curve, this is also noted. Values are bracketed where we cannot favour one interpretation over an other or the interpretation is not unequivocal. For those bursts with no valid blast wave interpretation, values are not included.} 	
  \label{breaks:blast-table} 	
  \begin{tabular}{l c c c c l } 
    \hline 
    GRB &   & $p$ &  $k$  & $q$ &  \\ 
    \hline
    050730	& ($\nu_{{\rm X}}< \nu_{{\rm c}}$) & (2.46 $\pm$ 0.06)  & (--) & (--) &	(jet)  \\ 
    050801	& $\nu_{{\rm X}}< \nu_{{\rm c}}$	 & 2.64 $\pm$ 0.20  & -0.7 $\pm$ 1.4 & 0.65 $\pm$ 0.15	& \\ 
    050802	& $\nu_{{\rm X}}< \nu_{{\rm c}}$	 & 2.62 $\pm$ 0.06  & 1.7 $\pm$ 0.3  & 0.90 $\pm$ 0.13 & \\ 
    050820A	& --	& --	& --  & --	&  --  \\ 
    051109A	& $\nu_{{\rm c}}< \nu_{{\rm X}}$	 & 2.08 $\pm$ 0.08  & -1.7 $\pm$ 0.8  & 0    & jet	   \\ 
    060124	& $\nu_{{\rm c}}< \nu_{{\rm X}}$	 & 2.02 $\pm$ 0.06  & 0.0 $\pm$ 0.4   & 0    & jet \\ 
    060206	& ($\nu_{{\rm c}}< \nu_{{\rm X}}$)	 & (2.34 $\pm$ 0.16)  & (--)  &  (0)   & (jet)	  \\ 
        	& ($\nu_{{\rm X}}< \nu_{{\rm c}}$)	 & (3.34 $\pm$ 0.16)  & (1.3 $\pm$ 0.7) & (0.8 $\pm$ 0.2) &  () \\ 
    060729	& $\nu_{{\rm c}}< \nu_{{\rm X}}$	 & 2.22 $\pm$ 0.04  & --             & 1.00 $\pm$ 0.03     & jet   \\
    061121	& $\nu_{{\rm c}}< \nu_{{\rm X}}$ 	 & 1.88 $\pm$ 0.08  & 1.5 $\pm$ 0.3  & 0 & jet \\ 
    061126	& --	& --	& --  &	-- &  -- \\ 
    \hline 
  \end{tabular}
\end{table*}

\subsubsection{GRB\,060729}

We find that $p=2.22 \pm 0.04$ for $\nu_{{\rm c}} < \nu_{{\rm X}}$ and $q = 1.00 \pm 0.03$ with $k$ unconstrained.
Despite the fact that the broadband SED seems to favour a broken power-law (though this is not conclusive), the parallel nature of the lightcurves suggests that there is no cooling break between regimes  so  both the X-ray and optical regimes are either above, or below, the cooling break. For $\nu_{{\rm X,opt}} > \nu_{{\rm c}}$, $p = 2.22 \pm 0.04$ implies $\alpha_{{\rm X,opt,}p} = 1.17 \pm 0.02$ in agreement with both the X-ray and optical temporal slopes, $\alpha_{{\rm X,2}}$ and $\alpha_{{\rm opt,2}}$. The break, to $\alpha = 1.70 \pm 0.09$, observed in the X-ray at $\sim 10^{6}$\,s is interpreted as a jet break, albeit a slightly shallow one, though it may be rolling over to a steeper asymptotic value. Here, the early shallow phase is likely due to continued energy injection of the form, $q = 1.00 \pm 0.03$, causing the observed X-ray break of $\Delta\alpha = 1.20 \pm 0.05$, and consistent with the optical. Such an energy injection would increase the energy of the afterglow over the observed period of injection by a factor of $f \sim 10$.

If $\nu_{{\rm X,opt}} < \nu_{{\rm c}}$,  $p = 3.22 \pm 0.04$ implies $\alpha_{p} = 1.67 \pm 0.02$ or $2.17 \pm 0.03$ for a constant density or wind driven medium, respectively, steeper than either observed slope before $\sim 10^{6}$\,s  but consistent with the slope after that time. The break with $\Delta\alpha \geq 0.39 \pm 0.04$, dependent on $k$, can be explained by energy injection of between $q = 0.25 \pm 0.03$ and $q = 0.37 \pm 0.04$.
This would then mean that the early shallow phase cannot be explained by energy injection of the same form and would require an additional energy injection component, so we favour our initial explanation.

\subsubsection{GRB\,061121}

We find that $p=1.88 \pm 0.08$ for $\nu_{{\rm c}} < \nu_{{\rm X}}$ and $k = 1.5 \pm 0.3$, $q=0$.
The broken power-law SED implies that the X-ray regime is above the cooling frequency, which in turn implies $p = 1.88 \pm 0.08$ and $\alpha_{{\rm X,}p} = 0.98 \pm 0.04$, in agreement with the observed X-ray index of $\alpha_{{\rm X,1}} = 0.97 \pm 0.02$ after 2000\,s. The optical temporal slope,  $\alpha_{{\rm opt,1}}$, is then explained as being below the cooling break and in a medium with $k=1.8 \pm 0.3$. The break observed in the X-ray light curve and consistent with the optical data is interpreted as a jet break rolling over to the expected value of $\alpha_{{\rm jet}} = p$, in agreement with the observed value of $\alpha_{{\rm X,2}} = 1.55 \pm 0.03$. It is unclear what the X-ray emission before 2000\,s is, but given that no shallowing is observed in optical, we can assume that it is not caused by continued energy injection, at least not in isolation. If it were due to continued energy injection, it would have a value of $q \sim 0.55$.

\subsubsection{GRB\,061126}

We do not find a blast wave interpretation consistent with the light curves of this burst. 
The inconsistency of the optical and X-ray slopes suggests that the cooling frequency is between the two regimes and $p = 1.82 \pm 0.08$, in agreement with the SED which is best described with a broken power-law. When  $\nu_{{\rm X}} > \nu_{{\rm c}}$, the predicted value of $\alpha_{{\rm X,}p} = 0.97 \pm 0.04$ significantly ($8\sigma$) underestimates the observed value of $\alpha_{{\rm X}} = 1.29 \pm 0.01$, though the optical can be explained in this scenario if the circumburst density profile has an index of  $k = 1.5 \pm 0.3$. 
The post-break optical slope of $\alpha_{{\rm opt}} = 1.9 \pm 0.1$ and an assumed `hidden' X-ray break could be consistent with the predicted $\alpha_{{\rm jet}} = p$ in this interpretation. However, given the disagreement of the pre-break X-ray temporal slope with the predicted value, it seems that the data of this burst are inconsistent with the blast wave model. This is consistent with \citet{gomboc2008:ApJ687} who suggest that the X-ray afterglow is due to an additional emission process rather than blast-wave shock emission.

\subsection{Distribution of $p$}\label{breaks:p-distribution}

\begin{figure} 
  \centering 
  \resizebox{\hsize}{!}{\includegraphics[angle=-90]{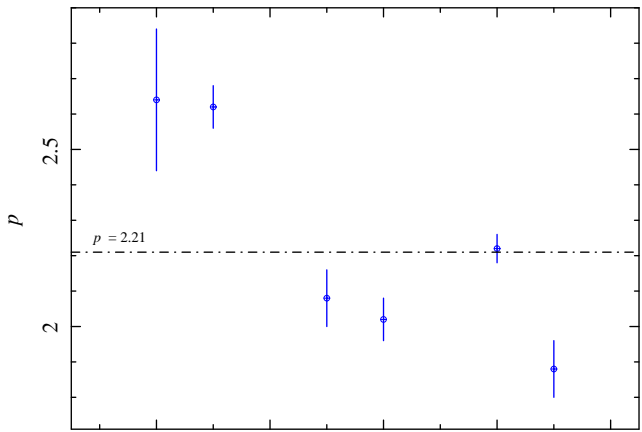} }
  \caption{Most likely values for $p$ after our interpretation of individual bursts (Table\,\ref{breaks:blast-table}); from left to right: GRB\,050801, GRB\,050802, GRB\,051109A, GRB\,060124, GRB\,060729, GRB\,061121. The line represents the most likely value of $p$ over the plotted sample  (Section \ref{breaks:p-distribution}).}
  \label{breaks:p-plot} 
\end{figure}

The universality of the electron energy distribution index, $p$, has been examined by several authors; \cite{chevalier2000:ApJ536}, \cite{panaitescu2002:ApJ571}, \cite{shen2006:MNRAS371} and \cite{starling2008:ApJ672} applied different methods to samples of \emph{BeppoSAX} and \emph{Swift} bursts, all reaching the same conclusion that the observed range of $p$ values is not consistent with a single central value of $p$. 
\cite{shen2006:MNRAS371} and \cite{starling2008:ApJ672} showed that the width of the parent distribution is $\sim 0.3-0.4$. In the studies so far there have been some limitations: the \emph{BeppoSAX} sample is limited, both in the number of GRBs and the temporal and spectral sampling; and the only study of \emph{Swift} bursts for this purpose so far by \citeauthor{shen2006:MNRAS371} only used the X-ray afterglows, which introduces a large uncertainty because the position of the cooling break is unknown. Here we examine the universatily of $p$ for our sample of  \emph{Swift} bursts by using the same methods as described in \cite{starling2008:ApJ672}. We test the null hypothesis that the observed distribution of $p$ (Table \ref{breaks:blast-table} excluding bracketed values and Figure \ref{breaks:p-plot}) can be obtained from a parent distribution with a single central value of $p$.

We minimise the log-likelihood as given in equation 3 of \citeauthor{starling2008:ApJ672}, and we find the most likely value of $p=2.21\pm 0.03$. To test our null hypothesis, we generate $10^{5}$ different synthetic data sets for $p$ for the 6 bursts in our sample by taking random numbers from probability distributions that are described by the most likely value of $p=2.21$ and the $1\,\sigma$ uncertainties in the 6 values of $p$. We then take 6 values of $p$ coming from these synthetic data sets and calculate the most likely value of $p$ and the accompanying log-likelihood. Comparing the log-likelihood of the synthetic data with that coming from the measured $p$ values, we conclude that the null hypothesis is rejected at the $>5\,\sigma$ level. Following \citeauthor{starling2008:ApJ672}, we can also derive the width of the parent distribution, which is $0.24<\sigma_{\rm{scat}}<0.35$ at the $1\,\sigma$ level and $0.12<\sigma_{\rm{scat}}<1.22$ at the $3\,\sigma$ level.
We also tested the 10 values of $p$ from Table \ref{breaks:p-table} that offered the least deviation from the expected canonical value of $\sim 2.2$, with no regard for the interpretation of the light curves or SEDs. In this case, the most likely value of $p$ is $2.19 \pm 0.02$ and the values are inconsistent with one population at the $5\sigma$ level.

This result confirms the results from previous studies and has important implications for theoretical particle acceleration studies. Some of these (semi-)analytical calculations and simulations indicate that there is a nearly universal value of $p \sim 2.2-2.3$ (e.g. \citealt{kirk2000:ApJ542,achterberg2001:MNRAS328}), while other studies suggest that there is a large range of possible values for $p$ of $1.5-4$ \citep{baring2004:NuPhS136}. 
Although we find that there is not a universal value of $p$, our values for the width of the parent distribution indicate that it is not as wide as the latter study suggest. The width we find, $\sigma_{\rm{scat}}\sim0.3$ is comparable to the numbers found by \citeauthor{shen2006:MNRAS371} and \citeauthor{starling2008:ApJ672}, but our results are based on a sample with better  temporal and spectral sampling per GRB, on average.

Given that the value of $p$ is not universal, it may be possible that it changes suddenly or evolves gradually with time or radius even in a single event as environmental blast wave parameters (e.g., magnetic field, ambient density) change or evolve (e.g., \citealt{hamilton1992:ApJ398,vlahos2004:ApJ608,kaiser2005:MNRAS360}). If this is the case our derived values of $p$ should be considered time averaged values of the parameter. A change or evolution of $p$ should be observable as a change or evolution of the synchrotron spectral index, $\beta (p)$, but should also cause a deviation from the power or broken power-law decay indices ($\alpha (p)$, Table\,\ref{breaks:p-relations}) of the light curves, though neither of these effects is observable in our data.

\subsection{Circumburst density profile, $k$}\label{breaks:medium}

The density structure, or profile, of the circumburst medium is generally given as $\rho$, or $n$ (number density), $\propto r^{-k}$ where $k=0$ is a constant density, or ISM-like, medium and $k=2$ is a wind-like medium. The value of $k$ has important implications for the study of progenitor models as the currently favoured model for long GRBs, involving the collapse of a massive, Wolf-Rayet star, is expected to have an associated strong stellar wind affecting the circumburst environment. Support for a stellar wind has been observed via absorption lines with relative velocities in UV spectra in a number of cases, most notably GRB\,021004 (e.g., \citealt{schaefer2003:ApJ588}; \citealt{starling2005:MNRAS360}). However, \cite{panaitescu2002:ApJ571} found, from broadband modeling of 10 bursts, that a constant density medium was favoured in half the cases and in only one case was a wind driven profile necessary. \cite{starling2008:ApJ672} found, from a sample of 10 bursts, that 4 were likely wind driven while 2 were certainly not consistent with a wind.

\begin{figure} 
  \centering 
  \resizebox{\hsize}{!}{\includegraphics[angle=-90]{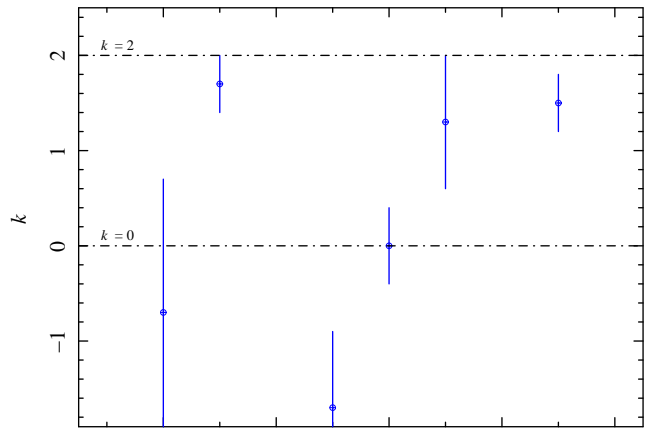} }
  \caption{Calculated values of $k$ after our interpretation of individual bursts (Table\,\ref{breaks:blast-table}); from left to right: GRB\,050801, GRB\,050802, GRB\,051109A, GRB\,060124, (GRB\,060206),  GRB\,061121. The lines represent $k=0$ (constant density) and $k=2$ (wind like) media (Section \ref{breaks:medium}). For bursts where we cannot favour one interpretation over an other or the interpretation is not unequivocal, we have bracketed the GRB name.}
  \label{breaks:k-plot} 
\end{figure}

In our sample of 10 bursts, only 6 have X-ray or optical light curves below the cooling break, $\nu_{{\rm c}}$, where they are dependent on the circumburst density profile and in the case of GRB\,060206 this interpretation is only one of two that cannot be distinguished.
Of our calculated values of $k$ (Table\,\ref{breaks:blast-table} and Figure\,\ref{breaks:k-plot}), 2 are consistent ($2\sigma$) with both $k=2$ and $k=0$ (GRB\,050801 and GRB\,060206); 2 are best described with a wind-like medium and are inconsistent ($5\sigma$) with $k=0$ (GRB\,050802 and GRB\,061121); and 2 are consistent ($2.1\sigma$) with a constant density medium but inconsistent with $k=2$  (GRB\,051109A and GRB\,060124). Our results are hence consistent with those of \citep{panaitescu2002:ApJ571} and \citep{starling2008:ApJ672} insofar as the sample requires both constant density and wind driven media to explain the observed broadband emission, as the sample is not consistent with only one of the two environments.

It seems clear that the circumburst environments are unlikely to be drawn from only one of the ISM or wind-like profiles and, in fact, the wind-like expectation is only valid assuming a constant mass-loss rate and wind velocity form the progenitor star; precarious assumptions for a star in its final death throws and given that the circumburst density profile may be affected by a number of processes. 
Firstly, a reverse shock is expected to propagate backwards through the wind as it meets the interstellar medium and would cause a constant density, shocked wind. 
Also, the unshocked ISM itself has an constant density profile but its distance from the burst makes it unlikely that we would observe the blast wave traversing the ISM. However, recent observations of Wolf-Rayet binaries suggests that this shock front might be much closer to the progenitor if the wind is asymmetric \citep{eldridge2007:MNRAS377}. This would make it more likely that the blast wave would already be travelling through the ISM at the time of observations, as would a high progenitor velocity towards the observer. In fact, the transition of the blast wave across the wind's shock front to the unshocked ISM has been suggested in a small number of cases, for instance, GRB\,030329 \citep{peer2006:ApJ643}. 
Not only can we expect density profiles of $k=2$ or $k=0$; \citeauthor{starling2008:ApJ672} discuss the fact that in Type Ib/c supernovae, with which GRBs are closely associated (e.g., GRB\,030329/SN\,2003dh, \citealt{hjorth2003:Natur423}), density profiles are found to range from $k=1.5$ to $k=2.0$ (see \citealt{weiler2004:NewAR48} and references therein).

\subsection{Rate of continued energy injection, $q$}\label{breaks:injection}

\cite{nousek2006:ApJ642} proposed continued energy injection as a process to explain the shallower than expected decay indices observed in many \emph{Swift} bursts. We assume this takes the form  of $E \propto t^{q}$ where $q$ is dependent on the shallowing effect of the injection, $\Delta\alpha = \alpha_{p} - \alpha_{{\rm observed}}$, and on the value of $p$. The continued energy injection is either due to $i)$ a distribution of the Lorentz factors of the shells ejected from the central engine which causes those shells with lower Lorentz factor to catch up with the blast wave at a later time, or $ii)$ continued activity of the central engine itself. The former has a limit of $q \leq 3-k$, where $k$ is the density profile of the circumburst medium, which \citeauthor{nousek2006:ApJ642} suggested could be used as a diagnostic to differentiate between the two sources of continued energy injection.

\begin{figure} 
  \centering 
  \resizebox{\hsize}{!}{\includegraphics[angle=-90]{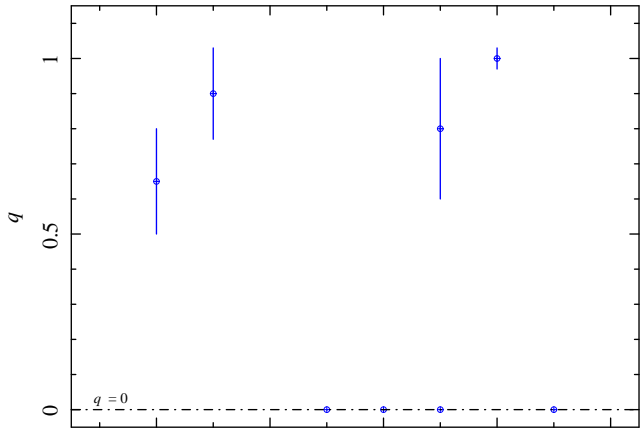} }
  \caption{Calculated values of $q$, including those with $q \equiv 0$ on line, after our interpretation of individual bursts (Table\,\ref{breaks:blast-table}); from left to right: GRB\,050801, GRB\,050802,  GRB\,051109A,  GRB\,060124, (GRB\,060206), GRB\,060729,  GRB\,061121 (Section \ref{breaks:injection}). For bursts where we cannot favour one interpretation over an other or the interpretation is not unequivocal, we have bracketed the GRB name.}
  \label{breaks:q-plot} 
\end{figure}

For 7 of the 10 light curves in our sample we are able to estimate the level of energy injection, or to say that it is consistent with zero (Table\,\ref{breaks:blast-table} and Figure\,\ref{breaks:q-plot}), though in the case of GRB\,060206 there are two equally valid interpretations, one requiring energy injection while the other does not. Excluding GRB\,060206, energy injection is only required in 3 out of 6 bursts and each of these has an injection index, $q$, of between 0.65 and 1.0, corresponding to increases in the energy of the afterglows over the observed periods of injection of a factor, $f$ of between 5 and 10. These values of $q$ and $f$ are consistent with both scenarios of energy injection and with previously calculated values of \cite{nousek2006:ApJ642}.

Clearly a much larger sample of multi-wavelength afterglows is required to be able to constrain the sample properties of $q$. These results compare to those of \cite{evans2009:mnras} who find that the majority of XRT light curves with observed breaks require energy injection. However, those authors study a significantly different, much larger sample of bursts while we consider a sample of only the most well sampled.

\section{Conclusion}\label{breaks:conclusion}

Throughout this paper we have applied the blast wave model \citep{rees1992:MNRAS258,meszaros1998:ApJ499}, assuming on-axis viewing, a standard jet structure and no evolution of the microphysical parameters, to a selection of GRBs with well sampled temporal and spectral data. We attempted to constrain three parameters of interest and to test the validity of the blast wave model for light curves observed by \emph{Swift}, the complexity of which has made the identification of breaks and the interpretation of the blast wave more difficult than in the pre-\emph{Swift} era. We find that the majority of the afterglows are well described within the frame work of the blast wave model and that the parameters derived are consistent with those values found by previous authors.

After an inspection of all the \emph{Swift} X-ray light curves and a comparison with the optical data available in the literature, we identified a sample of 10 bursts that had enough data to well constrain the optical and X-ray temporal indices, and the X-ray spectral indices, of the afterglows. We analysed these data, fitting power-law decays to the light curves in an attempt to identify possible breaks; taking into account the possibility of hidden breaks and the effect of host galaxy contribution to the optical, where possible.

In the case of well constrained X-ray spectra, we show that $p$ can be estimated from the X-ray spectra alone, with the caveat that it will give two possible values of $p$; these must be differentiated by either broadband SEDs or a comparison with the predictions of the blast wave model, or both.  
We compare the observed light curves and spectra of each burst in our sample to those values expected from the blast wave model and find that we can successfully interpret all bursts, except two, within this frame work; these interpretations require  `hidden' X-ray breaks in GRB\,060206 and GRB\,061126. Furthermore, we identify, reasonably unambiguously, jet breaks in 5 afterglows out of our sample of 10, including a jet break in GRB\,060729 that was not identified in previous analysis of the light curve. 

After interpretation within the blast wave model,  we are able to confidently estimate the electron energy distribution index, $p$, for 6 of the bursts in our sample. A statistical analysis of the distribution of $p$ reveals that, even in the most conservative case of least scatter, the values are not consistent with a single, universal value suggested by some studies; this has important implications for theoretical particle acceleration studies. 
In a number of cases, we are also able to obtain values for the circumburst density profile index, $k$, and the index of continued energy injection, $q$. The calculated values of $q$ are consistent with both suggested sources of continued energy injection and a much larger sample of afterglows will be required to constrain the sample properties of $q$. The values of $k$, consistent with previous works on the matter, suggest that the circumburst density profiles are not drawn from only one of the constant density or wind-like media populations.


\section*{Acknowledgements}

We thank P.A. Evans and K.L. Page for useful discussions on the XRT. 
We thank the referee for constructive comments.  
PAC \& RAMJW gratefully acknowledge support of NWO under Vici grant 639.043.302.
PAC \& RLCS  acknowledge support from STFC.
AJvdH was supported by an appointment to the NASA Postdoctoral Program at the MSFC, administered by Oak Ridge Associated Universities through a contract with NASA.
This work made use of data supplied by the UK Swift Science Data Centre at the University of Leicester funded by STFC and through the High Energy Astrophysics Science Archive Research Center Online Service, provided by the NASA/Goddard Space Flight Center.

\label{lastpage}


\begin{thebibliography}{}

\bibitem[\protect\citeauthoryear{{Achterberg} et~al.}{{Achterberg}
  et~al.}{2001}]{achterberg2001:MNRAS328}
{Achterberg}, A., {Gallant}, Y.~A., {Kirk}, J.~G.,  \& {Guthmann}, A.~W. 2001,
  \mnras, 328, 393

\bibitem[\protect\citeauthoryear{{Alatalo}, {Perley}, \& {Bloom}}{{Alatalo}
  et~al.}{2006}]{alatalo2006:GCN4702}
{Alatalo}, K., {Perley}, D.,  \& {Bloom}, J.~S. 2006, \gcn, 4702

\bibitem[\protect\citeauthoryear{{Baring}}{{Baring}}{2004}]{baring2004:NuPhS13%
6}
{Baring}, M.~G. 2004, Nuclear Physics B Proceedings Supplements, 136, 198

\bibitem[\protect\citeauthoryear{{Beuermann} et~al.}{{Beuermann}
  et~al.}{1999}]{beuermann1999:A&A352}
{Beuermann}, K., et~al. 1999, \aap, 352, L26

\bibitem[\protect\citeauthoryear{{Bloom}, {Perley}, \& {Chen}}{{Bloom}
  et~al.}{2006}]{bloom2006:GCN5826}
{Bloom}, J.~S., {Perley}, D.~A.,  \& {Chen}, H.~W. 2006, \gcn, 5826

\bibitem[\protect\citeauthoryear{{Burrows} et~al.}{{Burrows}
  et~al.}{2007}]{burrows2007:RSPT365}
{Burrows}, D.~N., et~al. 2007, Royal Society of London Philosophical
  Transactions Series A, 365, 1213

\bibitem[\protect\citeauthoryear{{Burrows} et~al.}{{Burrows}
  et~al.}{2005}]{burrows2005:SSRv120}
{Burrows}, D.~N., et~al. 2005, Space Science Reviews, 120, 165

\bibitem[\protect\citeauthoryear{{Burrows} \& {Racusin}}{{Burrows} \&
  {Racusin}}{2007}]{burrows2007:astro.ph2633}
{Burrows}, D.~N.,  \& {Racusin}, J. 2007, Il Nuovo Cimento B, 121, 1273

\bibitem[\protect\citeauthoryear{{Butler} \& {Kocevski}}{{Butler} \&
  {Kocevski}}{2007}]{butler2007:ApJ663}
{Butler}, N.~R.,  \& {Kocevski}, D. 2007, \apj, 663, 407

\bibitem[\protect\citeauthoryear{{Cenko}, {Berger}, \& {Cohen}}{{Cenko}
  et~al.}{2006}]{cenko2006:GCN4592}
{Cenko}, S.~B., {Berger}, E.,  \& {Cohen}, J. 2006, \gcn, 4592

\bibitem[\protect\citeauthoryear{{Cenko} et~al.}{{Cenko}
  et~al.}{2006}]{cenko2006:ApJ652}
{Cenko}, S.~B., et~al. 2006, \apj, 652, 490

\bibitem[\protect\citeauthoryear{{Chen} et~al.}{{Chen}
  et~al.}{2005}]{chen2005:GCN3709}
{Chen}, H.-W., {Thompson}, I., {Prochaska}, J.~X.,  \& {Bloom}, J. 2005, \gcn,
  3709

\bibitem[\protect\citeauthoryear{{Chevalier} \& {Li}}{{Chevalier} \&
  {Li}}{2000}]{chevalier2000:ApJ536}
{Chevalier}, R.~A.,  \& {Li}, Z.-Y. 2000, \apj, 536, 195

\bibitem[\protect\citeauthoryear{{Chincarini} et~al.}{{Chincarini}
  et~al.}{2007}]{chincarini2007:ApJ671}
{Chincarini}, G., et~al. 2007, \apj, 671, 1903

\bibitem[\protect\citeauthoryear{{Curran} et~al.}{{Curran}
  et~al.}{2006}]{curran2006:astro.ph.10067}
{Curran}, P.~A., {Kann}, D.~A., {Ferrero}, P., {Rol}, E.,  \& {Wijers},
  R.~A.~M.~J. 2006, Il Nuovo Cimento 121 B, 12, 1461

\bibitem[\protect\citeauthoryear{{Curran} et~al.}{{Curran}
  et~al.}{2008}]{curran2008:A&A487}
{Curran}, P.~A., {Starling}, R.~L.~C., {O'Brien}, P.~T., {Godet}, O., {van der
  Horst}, A.~J.,  \& {Wijers}, R.~A.~M.~J. 2008, \aap, 487, 533

\bibitem[\protect\citeauthoryear{{Curran} et~al.}{{Curran}
  et~al.}{2007a}]{curran2007:A&A467}
{Curran}, P.~A., et~al. 2007a, \aap, 467, 1049

\bibitem[\protect\citeauthoryear{{Curran}, {van der Horst}, \&
  {Wijers}}{{Curran} et~al.}{2008}]{curran2008:MNRAS386}
{Curran}, P.~A., {van der Horst}, A.~J.,  \& {Wijers}, R.~A.~M.~J. 2008,
  \mnras, 386, 859

\bibitem[\protect\citeauthoryear{{Curran} et~al.}{{Curran}
  et~al.}{2007b}]{curran2007:MNRAS381}
{Curran}, P.~A., et~al. 2007b, \mnras, 381, L65

\bibitem[\protect\citeauthoryear{{de Pasquale} et~al.}{{de Pasquale}
  et~al.}{2007}]{depasquale2007:MNRAS377}
{de Pasquale}, M., et~al. 2007, \mnras, 377, 1638

\bibitem[\protect\citeauthoryear{{de Pasquale} et~al.}{{de Pasquale}
  et~al.}{2006}]{depasquale2006:A&A455}
{de Pasquale}, M., et~al. 2006, \aap, 455, 813

\bibitem[\protect\citeauthoryear{{Eldridge}}{{Eldridge}}{2007}]{eldridge2007:M%
NRAS377}
{Eldridge}, J.~J. 2007, \mnras, 377, L29

\bibitem[\protect\citeauthoryear{{Evans} et~al.}{{Evans}
  et~al.}{2007}]{evans2007:A&A469}
{Evans}, P.~A., et~al. 2007, \aap, 469, 379

\bibitem[\protect\citeauthoryear{{Evans} et~al.}{{Evans}
  et~al.}{2008}]{evans2009:mnras}
{Evans}, P.~A., et~al. 2008, MNRAS, submitted (arXiv:0812.3662)

\bibitem[\protect\citeauthoryear{{Falcone} et~al.}{{Falcone}
  et~al.}{2007}]{falcone2007:ApJ671}
{Falcone}, A.~D., et~al. 2007, \apj, 671, 1921

\bibitem[\protect\citeauthoryear{{Fynbo} et~al.}{{Fynbo}
  et~al.}{2005}]{fynbo2005:GCN3749}
{Fynbo}, J.~P.~U., et~al. 2005, \gcn, 3749

\bibitem[\protect\citeauthoryear{{Fynbo} et~al.}{{Fynbo}
  et~al.}{2006}]{fynbo2006:A&A451}
{Fynbo}, J.~P.~U., et~al. 2006, \aap, 451, L47

\bibitem[\protect\citeauthoryear{{Gehrels} et~al.}{{Gehrels}
  et~al.}{2004}]{gehrels2004:ApJ611}
{Gehrels}, N., et~al. 2004, \apj, 611, 1005

\bibitem[\protect\citeauthoryear{{Gomboc} et~al.}{{Gomboc}
  et~al.}{2008}]{gomboc2008:ApJ687}
{Gomboc}, A., et~al. 2008, \apj, 687, 443

\bibitem[\protect\citeauthoryear{{Grupe} et~al.}{{Grupe}
  et~al.}{2007}]{grupe2007:ApJ662}
{Grupe}, D., et~al. 2007, \apj, 662, 443

\bibitem[\protect\citeauthoryear{{Halpern} \& {Armstrong}}{{Halpern} \&
  {Armstrong}}{2006a}]{halpern2006:GCN5853}
{Halpern}, J.,  \& {Armstrong}, E. 2006a, \gcn, 5853

\bibitem[\protect\citeauthoryear{{Halpern} \& {Armstrong}}{{Halpern} \&
  {Armstrong}}{2006b}]{halpern2006:GCN5851}
{Halpern}, J.,  \& {Armstrong}, E. 2006b, \gcn, 5851

\bibitem[\protect\citeauthoryear{{Halpern}, {Mirabal}, \&
  {Armstrong}}{{Halpern} et~al.}{2006a}]{halpern2006:GCN5847}
{Halpern}, J.~P., {Mirabal}, N.,  \& {Armstrong}, E. 2006a, \gcn, 5847

\bibitem[\protect\citeauthoryear{{Halpern}, {Mirabal}, \&
  {Armstrong}}{{Halpern} et~al.}{2006b}]{halpern2006:GCN5840}
{Halpern}, J.~P., {Mirabal}, N.,  \& {Armstrong}, E. 2006b, \gcn, 5840

\bibitem[\protect\citeauthoryear{{Hamilton} \& {Petrosian}}{{Hamilton} \&
  {Petrosian}}{1992}]{hamilton1992:ApJ398}
{Hamilton}, R.~J.,  \& {Petrosian}, V. 1992, \apj, 398, 350

\bibitem[\protect\citeauthoryear{{Hill} et~al.}{{Hill}
  et~al.}{2004}]{hill2004:SPIE5165}
{Hill}, J.~E., et~al. 2004, in X-Ray and Gamma-Ray Instrumentation for
  Astronomy XIII. Edited by Flanagan, Kathryn A.; Siegmund, Oswald H. W.
  Proceedings of the SPIE, Volume 5165, pp. 217-231 (2004)., ed. K.~A.
  {Flanagan} \& O.~H.~W. {Siegmund}, 217

\bibitem[\protect\citeauthoryear{{Hjorth} et~al.}{{Hjorth}
  et~al.}{2003}]{hjorth2003:Natur423}
{Hjorth}, J., et~al. 2003, \nat, 423, 847

\bibitem[\protect\citeauthoryear{{J{\'o}hannesson}, {Bj{\"o}rnsson}, \&
  {Gudmundsson}}{{J{\'o}hannesson} et~al.}{2006}]{johannesson2006:ApJ640}
{J{\'o}hannesson}, G., {Bj{\"o}rnsson}, G.,  \& {Gudmundsson}, E.~H. 2006,
  \apj, 640, L5

\bibitem[\protect\citeauthoryear{{Kaiser}}{{Kaiser}}{2005}]{kaiser2005:MNRAS36%
0}
{Kaiser}, C.~R. 2005, \mnras, 360, 176

\bibitem[\protect\citeauthoryear{{Kalberla} et~al.}{{Kalberla}
  et~al.}{2005}]{kalberla2005:A&A440}
{Kalberla}, P.~M.~W., {Burton}, W.~B., {Hartmann}, D., {Arnal}, E.~M.,
  {Bajaja}, E., {Morras}, R.,  \& {P{\"o}ppel}, W.~G.~L. 2005, \aap, 440, 775

\bibitem[\protect\citeauthoryear{{Kirk} et~al.}{{Kirk}
  et~al.}{2000}]{kirk2000:ApJ542}
{Kirk}, J.~G., {Guthmann}, A.~W., {Gallant}, Y.~A.,  \& {Achterberg}, A. 2000,
  \apj, 542, 235

\bibitem[\protect\citeauthoryear{{Ledoux} et~al.}{{Ledoux}
  et~al.}{2005}]{ledoux2005:GCN3860}
{Ledoux}, C., et~al. 2005, \gcn, 3860

\bibitem[\protect\citeauthoryear{{Liang} et~al.}{{Liang}
  et~al.}{2008}]{liang2008:ApJ675}
{Liang}, E.-W., {Racusin}, J.~L., {Zhang}, B., {Zhang}, B.-B.,  \& {Burrows},
  D.~N. 2008, \apj, 675, 528

\bibitem[\protect\citeauthoryear{{Liang}, {Zhang}, \& {Zhang}}{{Liang}
  et~al.}{2007}]{liang2007:ApJ670}
{Liang}, E.-W., {Zhang}, B.-B.,  \& {Zhang}, B. 2007, \apj, 670, 565

\bibitem[\protect\citeauthoryear{{Madau}}{{Madau}}{1995}]{madau1995:ApJ441}
{Madau}, P. 1995, {ApJ}, 441, 18

\bibitem[\protect\citeauthoryear{{Melandri} et~al.}{{Melandri}
  et~al.}{2006}]{melandri2006:GCN5827}
{Melandri}, A., et~al. 2006, \gcn, 5827

\bibitem[\protect\citeauthoryear{{M\'esz\'aros} \& {Rees}}{{M\'esz\'aros} \&
  {Rees}}{1997}]{meszaros1997:ApJ476}
{M\'esz\'aros}, P.,  \& {Rees}, M.~J. 1997, \apj, 476, 232

\bibitem[\protect\citeauthoryear{{M\'esz\'aros}, {Rees}, \&
  {Wijers}}{{M\'esz\'aros} et~al.}{1998}]{meszaros1998:ApJ499}
{M\'esz\'aros}, P., {Rees}, M.~J.,  \& {Wijers}, R.~A.~M.~J. 1998, {ApJ}, 499,
  301

\bibitem[\protect\citeauthoryear{{Misra} et~al.}{{Misra}
  et~al.}{2007}]{misra2007:A&A464}
{Misra}, K., {Bhattacharya}, D., {Sahu}, D.~K., {Sagar}, R., {Anupama}, G.~C.,
  {Castro-Tirado}, A.~J., {Guziy}, S.~S.,  \& {Bhatt}, B.~C. 2007, \aap, 464,
  903

\bibitem[\protect\citeauthoryear{{Nousek} et~al.}{{Nousek}
  et~al.}{2006}]{nousek2006:ApJ642}
{Nousek}, J.~A., et~al. 2006, \apj, 642, 389

\bibitem[\protect\citeauthoryear{{Oates} et~al.}{{Oates}
  et~al.}{2007}]{oates2007:MNRAS380}
{Oates}, S.~R., et~al. 2007, \mnras, 380, 270

\bibitem[\protect\citeauthoryear{{O'Brien} et~al.}{{O'Brien}
  et~al.}{2006}]{obrien2006:ApJ647}
{O'Brien}, P.~T., et~al. 2006, \apj, 647, 1213

\bibitem[\protect\citeauthoryear{{Page} et~al.}{{Page}
  et~al.}{2007}]{page2007:ApJ663}
{Page}, K.~L., et~al. 2007, \apj, 663, 1125

\bibitem[\protect\citeauthoryear{{Panaitescu} \& {Kumar}}{{Panaitescu} \&
  {Kumar}}{2002}]{panaitescu2002:ApJ571}
{Panaitescu}, A.,  \& {Kumar}, P. 2002, \apj, 571, 779

\bibitem[\protect\citeauthoryear{{Panaitescu} et~al.}{{Panaitescu}
  et~al.}{2006}]{panaitescu2006:MNRAS369}
{Panaitescu}, A., {M{\'e}sz{\'a}ros}, P., {Burrows}, D., {Nousek}, J.,
  {Gehrels}, N., {O'Brien}, P.,  \& {Willingale}, R. 2006, {MNRAS}, 369, 2059

\bibitem[\protect\citeauthoryear{{Pandey} et~al.}{{Pandey}
  et~al.}{2006}]{pandey2006:A&A460}
{Pandey}, S.~B., et~al. 2006, \aap, 460, 415

\bibitem[\protect\citeauthoryear{{Pe'er} \& {Wijers}}{{Pe'er} \&
  {Wijers}}{2006}]{peer2006:ApJ643}
{Pe'er}, A.,  \& {Wijers}, R.~A.~M.~J. 2006, \apj, 643, 1036

\bibitem[\protect\citeauthoryear{{Perley} et~al.}{{Perley}
  et~al.}{2008}]{perley2008:ApJ672}
{Perley}, D.~A., et~al. 2008, \apj, 672, 449

\bibitem[\protect\citeauthoryear{{Quimby} et~al.}{{Quimby}
  et~al.}{2005}]{quimby2005:GCN4221}
{Quimby}, R., {Fox}, D., {Hoeflich}, P., {Roman}, B.,  \& {Wheeler}, J.~C.
  2005, \gcn, 4221

\bibitem[\protect\citeauthoryear{{Racusin} et~al.}{{Racusin}
  et~al.}{2008}]{racusin2008:Natur455}
{Racusin}, J.~L., et~al. 2008, \nat, 455, 183

\bibitem[\protect\citeauthoryear{{Rees} \& {M\'esz\'aros}}{{Rees} \&
  {M\'esz\'aros}}{1992}]{rees1992:MNRAS258}
{Rees}, M.~J.,  \& {M\'esz\'aros}, P. 1992, {MNRAS}, 258, 41P

\bibitem[\protect\citeauthoryear{{Rhoads}}{{Rhoads}}{1997}]{rhoads1997:ApJ487}
{Rhoads}, J.~E. 1997, {ApJ}, 487, L1

\bibitem[\protect\citeauthoryear{{Rykoff} et~al.}{{Rykoff}
  et~al.}{2006}]{rykoff2006:ApJ638}
{Rykoff}, E.~S., et~al. 2006, \apjl, 638, L5

\bibitem[\protect\citeauthoryear{{Sari}, {Piran}, \& {Narayan}}{{Sari}
  et~al.}{1998}]{sari1998:ApJ497}
{Sari}, R., {Piran}, T.,  \& {Narayan}, R. 1998, \apjl, 497, L17

\bibitem[\protect\citeauthoryear{{Schaefer} et~al.}{{Schaefer}
  et~al.}{2003}]{schaefer2003:ApJ588}
{Schaefer}, B.~E., et~al. 2003, \apj, 588, 387

\bibitem[\protect\citeauthoryear{{Schlegel}, {Finkbeiner}, \&
  {Davis}}{{Schlegel} et~al.}{1998}]{schlegel1998:ApJ500}
{Schlegel}, D.~J., {Finkbeiner}, D.~P.,  \& {Davis}, M. 1998, \apj, 500, 525

\bibitem[\protect\citeauthoryear{{Shen}, {Kumar}, \& {Robinson}}{{Shen}
  et~al.}{2006}]{shen2006:MNRAS371}
{Shen}, R., {Kumar}, P.,  \& {Robinson}, E.~L. 2006, \mnras, 371, 1441

\bibitem[\protect\citeauthoryear{{Stanek} et~al.}{{Stanek}
  et~al.}{2007}]{stanek2007:ApJ654}
{Stanek}, K.~Z., et~al. 2007, \apj, 654, L21

\bibitem[\protect\citeauthoryear{{Starling} et~al.}{{Starling}
  et~al.}{2008}]{starling2008:ApJ672}
{Starling}, R.~L.~C., {van der Horst}, A.~J., {Rol}, E., {Wijers}, R.~A.~M.~J.,
  {Kouveliotou}, C., {Wiersema}, K., {Curran}, P.~A.,  \& {Weltevrede}, P.
  2008, \apj, 672, 433

\bibitem[\protect\citeauthoryear{{Starling} et~al.}{{Starling}
  et~al.}{2005}]{starling2005:MNRAS360}
{Starling}, R.~L.~C., {Wijers}, R.~A.~M.~J., {Hughes}, M.~A., {Tanvir}, N.~R.,
  {Vreeswijk}, P.~M., {Rol}, E.,  \& {Salamanca}, I. 2005, \mnras, 360, 305

\bibitem[\protect\citeauthoryear{{Starling} et~al.}{{Starling}
  et~al.}{2007}]{starling2007:ApJ661}
{Starling}, R.~L.~C., {Wijers}, R.~A.~M.~J., {Wiersema}, K., {Rol}, E.,
  {Curran}, P.~A., {Kouveliotou}, C., {Van der Horst}, A.~J.,  \& {Heemskerk},
  M.~H.~M. 2007, \apj, 661, 787

\bibitem[\protect\citeauthoryear{{Tagliaferri} et~al.}{{Tagliaferri}
  et~al.}{2005}]{tagliaferri2005:Natur436}
{Tagliaferri}, G., et~al. 2005, \nat, 436, 985

\bibitem[\protect\citeauthoryear{{Th{\"o}ne} et~al.}{{Th{\"o}ne}
  et~al.}{2006}]{thoene2006:GCN5373}
{Th{\"o}ne}, C.~C., et~al. 2006, \gcn, 5373

\bibitem[\protect\citeauthoryear{{Vlahos}, {Isliker}, \& {Lepreti}}{{Vlahos}
  et~al.}{2004}]{vlahos2004:ApJ608}
{Vlahos}, L., {Isliker}, H.,  \& {Lepreti}, F. 2004, \apj, 608, 540

\bibitem[\protect\citeauthoryear{{Weiler} et~al.}{{Weiler}
  et~al.}{2004}]{weiler2004:NewAR48}
{Weiler}, K.~W., {van Dyk}, S.~D., {Sramek}, R.~A.,  \& {Panagia}, N. 2004, New
  Astronomy Review, 48, 1377

\bibitem[\protect\citeauthoryear{{Wo{\'z}niak} et~al.}{{Wo{\'z}niak}
  et~al.}{2006}]{wozniak2006:ApJ642}
{Wo{\'z}niak}, P.~R., {Vestrand}, W.~T., {Wren}, J.~A., {White}, R.~R.,
  {Evans}, S.~M.,  \& {Casperson}, D. 2006, \apj, 642, L99

\bibitem[\protect\citeauthoryear{{Yost}, {Schaefer}, \& {Yuan}}{{Yost}
  et~al.}{2006}]{yost2006:GCN5824}
{Yost}, S.~A., {Schaefer}, B.~E.,  \& {Yuan}, F. 2006, \gcn, 5824

\bibitem[\protect\citeauthoryear{{Yost} et~al.}{{Yost}
  et~al.}{2007}]{yost2007:ApJ657}
{Yost}, S.~A., et~al. 2007, \apj, 657, 925

\end{thebibliography}
\end{document}